# Surrogate modeling and control of medical digital twins

Luis L. Fonseca[1][*][¶], Lucas Böttcher[1,2][¶], Borna Mehrad[1], Reinhard C. Laubenbacher[1]

1 Laboratory for Systems Medicine, Department of Medicine, University of Florida, Gainesville, FL, USA.

2 Department of Computational Science and Philosophy, Frankfurt School of Finance and Management, 60322 Frankfurt am Main, Germany.

* Corresponding author:

Email: llfonseca@medicine.ufl.edu (LLF)

¶ These authors contributed equally to this work.

# Abstract

The vision of personalized medicine is to identify interventions that maintain or restore a person's health based on their individual biology. Medical digital twins, computational models that integrate a wide range of health-related data about a person and can be dynamically updated, are a key technology that can help guide medical decisions. Such medical digital twin models can be high-dimensional, multi-scale, and stochastic. To be practical for healthcare applications, they often need to be simplified into low-dimensional surrogate models that can be used for optimal design of interventions. This paper introduces surrogate modeling algorithms for the purpose of optimal control applications. As a use case, we focus on agent-based models (ABMs), a common model type in biomedicine for which there are no readily available optimal control algorithms. By deriving surrogate models that are based on systems of ordinary differential equations, we show how optimal control methods can be employed to compute effective interventions, which can then be lifted back to a given ABM. The relevance of the methods introduced here extends beyond medical digital twins to other complex dynamical systems.

# Author Summary

Our work enables the efficient control of biomedical and related digital twins that are based on agent-based models (ABMs), with potential applications to the optimization of therapeutic interventions. Agent-based models, also known as individual-based models, are a class of computational stochastic models in which an agent's behavior is modeled by rules rather than by equations. This property makes them well suited for modeling biological systems for which detailed *in vivo* kinetic information is unavailable, such as many aspects of the human immune system. However, since ABMs are not equation-based, conventional optimization approaches, sensitivity analyses, and optimal control methods are inapplicable or computationally infeasible. One way to overcome this issue is to approximate a given ABM using an ordinary differential equation (ODE) surrogate model for this purpose. In this work, we propose several types of ODE surrogate models and show how to employ them to approximate ABMs for the purpose of solving optimal control problems. Most treatment optimization tasks in medicine can be formulated as optimal control problems. Hence, the approaches presented here will likely be useful in applications involving ABM-based medical digital twins.

# Introduction

The promise of personalized medicine is to determine interventions for maintaining or restoring the health of a person based on their individual biology. A key technology to realize this promise is the use of high-fidelity computational models that are calibrated to an individual patient and help guide optimal interventions to restore or maintain health. If the models can be dynamically updated to reflect changes in a person's health status, they fall into the category of *medical digital twins*. While there are some examples of currently used medical digital twins (see, e.g., [1]), there are many obstacles to be overcome to their development at scale.

Medical digital twins often need to capture mechanisms across different spatial and temporal scales; for instance, the mode of action of many drugs is intracellular, while their effects manifest themselves at the organ or organism scale. Depending on the application, e.g., immune system applications, digital twins may also need to account for stochastic effects. As a result, high-fidelity medical digital twins can require high-dimensional, multi-scale, hybrid, stochastic, dynamic computational models at their core. This requirement poses two problems: First, optimal control theory methods, well-developed for ordinary differential equation models, are not generally applicable to complex hybrid models, leaving only *ad hoc* methods for the identification of optimal interventions. Second, high-dimensional complex models of this kind pose computational challenges to simulation, optimization, and control tasks, and they do not readily lend themselves to available data assimilation methods for dynamic calibration to new patient data. These problems have been encountered in digital twin research beyond medicine, and there are no general solutions available at this time, as detailed in the 2023 report *Fundamental Research Gaps for Digital Twins*, prepared by the National Academies of Engineering, Science, and Medicine [2].

While physics-based systems of equations provide useful descriptions of many natural phenomena, certain aspects of human biology, such as the immune system, are more effectively represented by alternative model types like agent-based models (ABMs) [3,4]. These models, characterized by rule-based dynamics, are often employed to simulate complex and spatially heterogeneous processes. To address the challenges described above, we propose a surrogate modeling solution for control problems associated with ABMs. The basic idea is as follows. *For a given control problem, we construct a surrogate model consisting of a system of ordinary differential equations (ODEs) for which optimal control methods are readily applicable. This surrogate model can then be used as the computational core of a digital twin in place of the original problem.* This strategy was proposed initially in [5]. Here, we further develop and formalize this control approach, and present its first empirical application to control ABMs. This paper does not address the problem of personalizing a model to a given patient using patient-derived time course data. Even for ODE models, this poses formidable challenges that are beyond the scope of this study.

## Control

The concept of control is central to medicine, for example in containing the spread of a pathogen using antibiotics, regulating blood sugar levels in diabetic patients, or managing high blood pressure [6–8]. In such settings, it is often required to not only determine how to achieve control (*i.e.*, treat a specific condition), but also to do so optimally without over- or undershooting, and effectively minimizing treatment side effects and costs while ensuring the desired outcomes. To illustrate the need for techniques such as those presented here, in the context of medical digital twin applications, is the treatment of patients suffering from pneumonia in an Intensive Care Unit (ICU). A computational model underlying a pneumonia digital twin would capture the immune response to the pathogen in the lungs, e.g., a more comprehensive version of the model in [9]. It is a stochastic multiscale model, with an ABM at the tissue level, Boolean network models at the intracellular scale, and partial differential equations (PDEs) for diffusion of substances. This model could be personalized to the specific pneumonia patient using data assimilation methods, applied to a mixture of data collected from the patient over time and outcome data from a suitably chosen reference patient cohort. This personalized model can then be used to simulate the effect of different possible interventions at a given time point, such as the determination and duration of administration of antibiotics, maybe in combination with other drugs. The techniques in this paper can then be used to develop appropriate surrogate models for treatment optimization.

Most engineered systems can be described using physics-based ODE models, allowing for the application of well-established optimal control methods[6,10]. However, biomedical control problems present unique hurdles to this approach. Human biology exhibits significantly greater variability across individuals [11–14] compared to engineered systems. Additionally, many systems crucial for human health, such as the immune system or the human microbiome, encompass highly diverse cell populations [15]. These systems are characterized by high-dimensional dynamic processes that go beyond the scope of representation achievable with a relatively small number of ODEs, as is typical for engineered systems. The development of mathematical and computational models of complex biomedical dynamics is still in its infancy and demands further research [9,16–21]. In this paper, we introduce an efficient approach to model-based control that addresses some of these challenges. While our primary focus is on applications to medical digital twins in particular, the tools we present have broad applicability, extending, for instance, to ecological models like predator-prey systems [22–24].

## Agent-based models

A key modeling paradigm for parts of human biology that are not easily described through an equation-based approach is the class of *agent-based models* (ABMs), or *individual-based models* [9,19–21,25]. These are computational models that allow the explicit representation of entities (*i.e.*, agents) such as cells in a tissue or microbes in a biofilm [26–31]. Agents follow a set of typically stochastic rules to navigate a heterogeneous spatial environment and engage in interactions with one another. This class of models is intuitive and easy to implement and is hence well-suited to be used by domain experts without extensive knowledge of computational modeling. ABMs have been used in a wide range of medical applications, from studies of the immune system, tumor growth and treatment to drug development [9,19–21]. A major drawback of ABMs, however, is that they are typically not equation-based and are therefore not amenable to standard mathematical analysis, leaving simulation as the primary approach. Moreover, there is a lack of well-established mathematical tools for model-based control.

An *et al.* [5] proposed a general approach to finding effective controls for ABMs by approximating them with a system of ODEs. This approximation is designed to capture the fundamental aspects of the control problem, even if it might not precisely match an ABM's dynamics under different initial conditions. In essence, this method treats an ABM as a high-fidelity model of the real world, while the employed ODE system acts as a surrogate model specific to the control problem at hand. In this way, one can build a comprehensive high-resolution model of human biology, and then create a simpler surrogate model for a given control problem. For digital twin applications, this approach is more appropriate than creating individual models for each new control application. It also avoids the drawback of having to decide *a priori* what part of a patient's biology is relevant to the current control application.

We sought to implement this strategy and provide a detailed application in two specific use cases. In doing so, we combine the strengths of both ABM and ODE models: the intuitive computational modeling of spatially heterogeneous, rule-driven stochastic systems with the effectiveness of mathematical ODE control approaches. We summarize the different steps of the proposed control approach in Fig 1. Our method's broad applicability stems from its ability to (i) leverage the rich repertoire of modeling features of an ABM, (ii) closely approximate mechanistic features, when available, ensuring alignment with the underlying model mechanisms, and (iii) employ techniques to approximate an ABM that is near a relevant steady state for control purposes. While the ABMs we consider here are relatively simple compared to many ABMs appearing in realistic applications, they describe fundamental multi-species and biochemical dynamics that can be expected to arise in many applications of medical digital twins. Thus, they present many of the fundamental challenges to be addressed without obscuring them with biological and computational complexity. The steps described in the following sections can be applied to digital twin systems or any other stochastic model for which only forward simulations are feasible.

Several fundamental differences between ABMs and ODEs need to be considered. First, almost all ABMs are stochastic models, whereas ODE models are deterministic. Second, ABMs usually account for heterogeneous spatial features, including structured environments that agents navigate and a heterogeneous population of agents within this environment. As a result, an ODE approximation of an ABM is unlikely to fully capture ABM dynamics to a degree that it can serve as a universally reliable approximation for general model simulation. In essence, ABMs are characterized by microstates at the agent level that generate emergent dynamics of population-level macrostates, whereas ODE models can only capture aggregate macrostates. ODE approximations can, nevertheless, approximate the solution of a given control problem within a restricted domain. Additionally, previous studies have demonstrated that ODE surrogate models of complex networked dynamical systems can provide accurate descriptions of their evolution in certain cases [32,33].

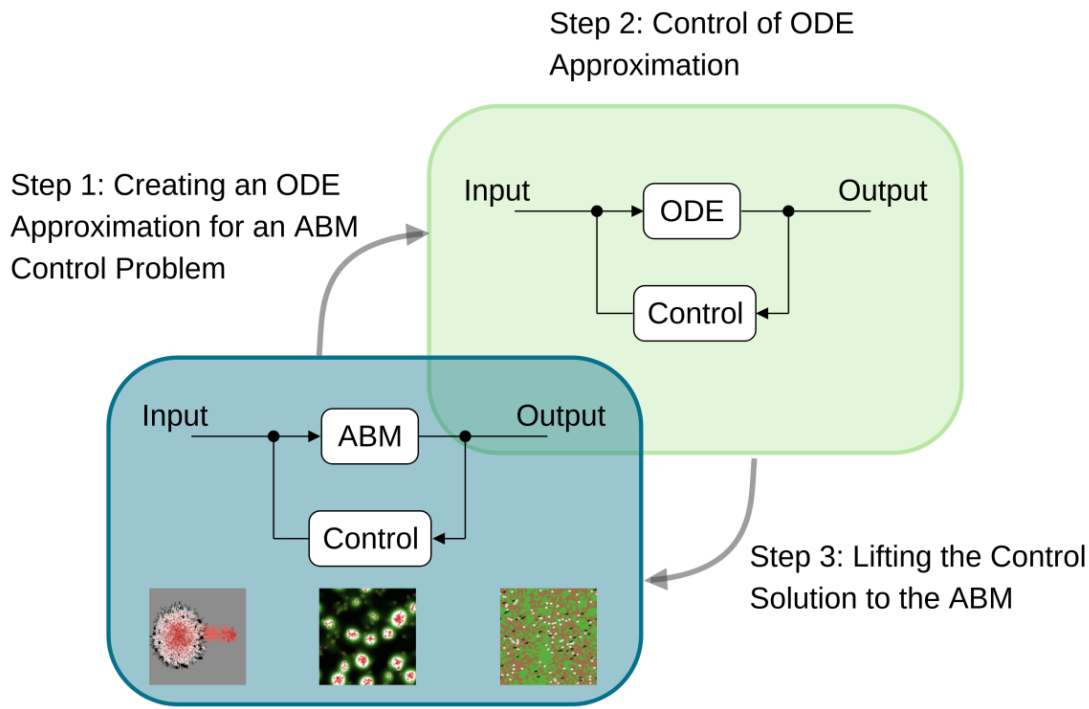

**Fig 1. Summary of the key steps involved in using ODE surrogate models for control.**
For an ABM associated with a control problem, we first create an ODE surrogate (Step 1). Next, we apply control techniques to the ODE system (Step 2) before lifting the control solution back to the original ABM (Step 3). The snapshots in the ABM panel depict simulations performed in NetLogo[34] of tumor[35], slime mold[36], and wolf sheep predation models[37].

The diversity in structure among ABMs poses a significant challenge for any general approximation method, including ours, rendering the formulation and implementation of a general algorithm difficult. The fact that there are no generally accepted implementation standards available for ABMs, contrary to the Systems Biology Markup Language (SBML) standard for ODEs [38–41], complicates this approach further. While the text-based Overview, Design concepts, Details (ODD) protocol [42,43] provides a useful framework for describing ABMs, it leaves many ambiguities in how to map such a description to an actual implementation [44].

## Surrogate Modeling

Surrogate models, also known as metamodels, proxy models, or emulators[45,46], are a class of models developed to approximate another model, often called the high-fidelity model. The former are usually simplified models, which allow one to predict the evolution of the latter using fewer computational resources, resulting in a tradeoff between accuracy and speed. Additionally, surrogate models are also commonly employed for performing sensitivity analysis, uncertainty quantification, risk analysis, and optimization. As pointed out by Bahrami *et al.* [45], surrogate modeling does not replace the development of high-fidelity, high-accuracy models. The latter are an important part of modeling large, complex, multiscale systems. Surrogate models serve mainly as supplementary tools for analyzing existing high-fidelity, high-accuracy models. They enable specific analyses that would otherwise be impractical (*e.g.*, optimal control) or resource-intensive (*e.g.*, sensitivity analysis and optimization).

In their review Bahrami *et al.* [45] describe a classification scheme of proxy models, which is valuable for characterizing the surrogate models developed here. In order to enable the application of optimal control methods within a dynamic ABM, two key requirements must be met: (i) the presence of a dynamic surrogate model, and (ii) compatibility with a formalism supporting optimal control theory. These criteria are both satisfied by employing an ODE-based surrogate model. Some authors [45,47] categorize surrogate models as either physics-based or black-box. For the purpose of this paper, we suggest using a spectrum of surrogate models. If enough information about a high-fidelity ABM is available, one can employ a mechanistic (physics-based) approximation. If limited information is available about a given ABM, or if an automatically generated surrogate model is preferred, we propose a technique for constructing a black-box model, such as an S-system[48,49]. However, if it is feasible to infer the flow of mass through the components of the high-fidelity ABM, one may consider using an ODE defined with a stoichiometric matrix that regulates this mass flow. Alternatively, the processes within the ODE could be approximated using general functions, such as power laws. This approach sidesteps the more challenging task of mechanistically approximating the processes. Hence, the resulting surrogate model is only semi-mechanistic, or what is sometimes referred to as a grey-box model.

After formulating the surrogate models, we propose to parameterize them through fitting, using maximum likelihood estimation against datasets produced by simulating the high-fidelity model. While various methods exist for dataset generation and fitting [47], we recommend selecting one based on the specific problem. We chose several datasets generated with diverse initial conditions and varying levels of control. This strategy ensures that the surrogate model effectively approximates the original model across a broad range of values in both the state variables and the control parameters.

After describing the algorithm in the Materials and Methods, we apply it to two ABMs. The first is a well-studied multi-species ABM from ecology, a predator-prey model with a resource component, implemented as wolves and sheep, with grass as the resource [37]. This model is implemented in the popular modeling platform Netlogo [34]. We have chosen this application for two reasons: (1) to demonstrate the power and versatility of our method without the technical challenges of a more complex ABM, and (2) because this model type is broadly used in biomedicine; for instance, most in-host infection models are essentially predator-prey models, where immune cells behave as predators and pathogenic cells as prey. The second ABM is a model of a metabolic network with 5 metabolites, 4 reactions and two regulatory interactions. This second model is larger and more complex than the first. It has both microstates and macrostates. The microstates are characterized by the elementary reactions between metabolites and enzymatic complexes, and the macrostates are given by Michaelis–Menten approximations. Many biological and medical processes are modeled using Michaelian or Monod processes. This second model was created with the objective of having a challenging test case for power law–based surrogate models, both in structure and initial conditions. We considered this step necessary to pinpoint the limitations of these surrogate models and explore potential alternative solutions.

# MATERIALS AND METHODS.

**Algorithm Input.** The starting point of the algorithm is a control problem derived from a medical application, using a digital twin with an underlying ABM. The ABM may be specified at different levels of detail, ranging from a detailed description and source code following an ODD protocol [42,43,50] to just an executable file. Relevant control problems include scenarios where a single control input is needed to transition the model from one steady state to another, as well as situations requiring feedback control to keep one or more state variables within a specified range.

## Mapping of ABM features to ODE features.

Rather than aiming to replicate all key ABM features in an ODE approximation, a more effective strategy is often to capture only the most critical features of an ABM relevant to the specific control problem. This process leads to a reduced ODE surrogate model that can be used for optimal control purposes. As mentioned earlier and further detailed in the Discussion section, there are several reasons why it might be useful to initially construct a complex high-fidelity ABM and then proceed to model this ABM with a coarser surrogate ODE model, rather than creating a simplified ODE model directly from the beginning. The detailed steps in this surrogate modeling process depend to some extent on the nature of the ABM. In many cases, it might be possible to use sensitivity analysis to determine the ABM parameters and state variables most important to the control problem and represent them explicitly in the ODE, while aggregating or eliminating the others.

Ordinary differential equations and ABMs are distinct modeling frameworks. We therefore first establish a correspondence between their respective model components (Fig 2).

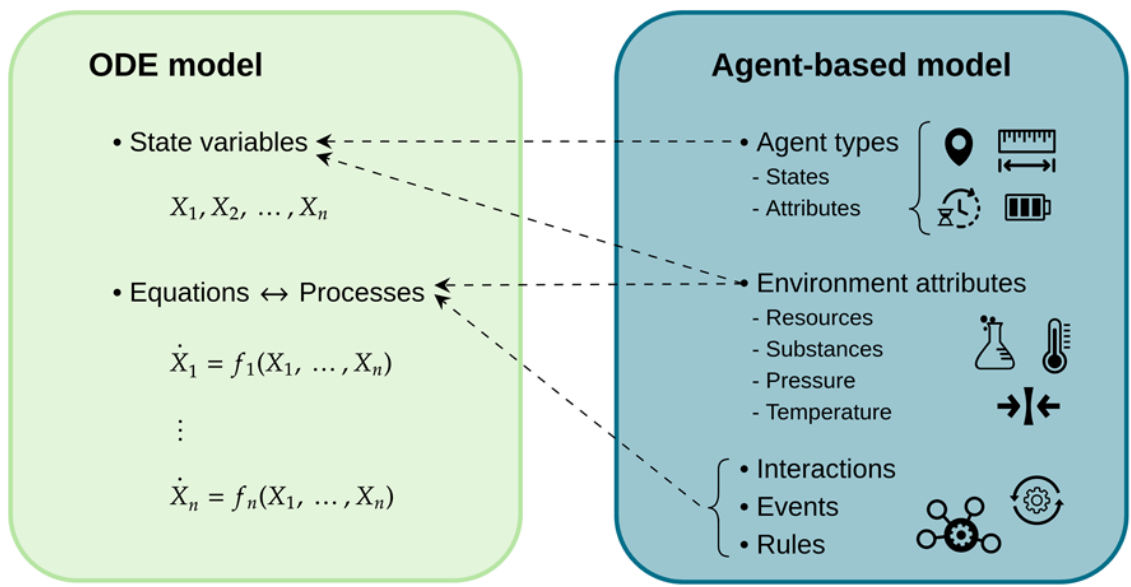

**Fig 2. Correspondence between ABM and ODE model components.** The aggregation of agents by type or attributes characterizes the state variables of an ODE approximation. Similarly, all interactions, events, and rules in the ABM characterize the processes of an ODE model. Depending on the ABM's structure, its environment can be transformed into state variables or contribute to the processes of an ODE approximation.

Ordinary differential equations are defined by a set of state variables, each of which changes according to an equation that sums up the different processes that affect the state variable in question. On the other hand, ABMs consist of several different components. The entities that collectively make up the state of an ABM include "agents" (or "individuals") that can take on values from a set of internal states and are equipped with a set of rules that govern their interactions with each other and with the environment in which they operate, resulting in changes of their internal states and spatial position. Other components may include properties of the spatial environment. Additionally, ABMs encompass a description of processes governed by agent and environment rules, along with scheduling details for these processes and information regarding their respective time scales. The ODD protocol [42,43,50] provides a systematic way to organize this information.

As the first step in the algorithm, we map all ABM components determined to be relevant to the optimal control problem to an ODE framework (Fig 2). This is a key step, and there are usually many choices of how to do this. Recall that the goal of the proposed approximation method is to solve a given ABM control problem by mapping it to an ODE model, use the rich set of control approaches available for such models, and then map the solution back to the ABM. The goal is not to identify the best ODE representation capable of reproducing the general dynamics of an ABM, which might require very different modeling choices.

The algorithm is divided into cases according to the information available about the ABM. The more information that can be used, the better the ODE approximation will generally be for solving the control problem. The best scenario is complete knowledge of the model components, that is, an ODD protocol for the model, which allows the construction of an ODE model based on key mechanistic features. At the other end of the spectrum is the ABM model as an executable program from which simulation trajectories can be obtained, but no mechanistic information is available.

**CASE 1: Information about ABM mechanisms is available.**

**Input:** Information from an ABM that enables the determination of the exact functional form and all terms of each differential equation (including the underlying stoichiometric matrix) in an ODE surrogate model.

If enough mechanistic information about an ABM is available, then it is possible to construct an ODE approximation that captures key mechanistic details relevant to the control problem, potentially leading to a better-performing approximation. For such a mechanistic approximation, we require information about the processes in the ABM to determine the complete model structure, (i) the terms appearing in each differential equation describing a state variable, and (ii) the exact functional form of each of these terms. Having access to a detailed description of the rules of an ABM allows us to determine the terms in the differential equations describing the evolution of each state variable in an ODE surrogate model. In the case of a biochemical network model, this would be the stoichiometric matrix of the model.

**Step 1:** _Identify the state variables of the ODE_. Any agent state that separately from all others modulates a process of interest or is changed by a process of interest, must be represented by its own state variable. Additionally, if control will be applied to a given agent type or state, then this agent will also have to be explicitly included as a state variable. Similarly, any agent type or state that will be used as a control signal independently of all others will have to be explicitly included as a state variable. If several agents always modulate the same processes in the same way, then it might be possible to aggregate or lump these agents into a single state variable. For example, if a certain agent type modulates a process independent of its position, then position-related information may be discarded, otherwise agents may need to be aggregated into position-dependent states.

**Step 2:** _Determine the stoichiometric matrix_. This requires identifying which state variables are changed by each of the processes (interactions or events) and specifying the sign of the effect: positive if the process leads to an increase of the state variable and negative if it leads to a decrease.

Once all state variables and processes have been identified, as well as the dependencies of variables on processes, then the system may be written as

$$\frac{dX}{dt} = M \cdot F \tag{1}$$

where $X = (X_1, \ldots , X_n)$ is the vector of all $n$ state variables, $F = (F_1, \ldots , F_m)$ is the vector of all $m$ processes, and $M = (a_{ij})$ is the $n \times m$ stoichiometric matrix. The stoichiometric matrix is usually sparse, with a non-zero entry $a_{ij}$ of either 1 or -1 in position $(i,j)$ indicating that the process $F_j$ affects the state variable $X_i$ positively or negatively, respectively.

**Step 3.** _Determine the functional form of the ODE process terms_. This is achieved by analyzing the nature of the processes underlying the interactions or rules of an ABM. In many cases, the processes have well-understood characteristics, such as mass action kinetics [51–55], typical of biochemical systems, epidemiological systems like SIR [56], ecological systems like Lotka–Volterra models [57,58] and related population-based models; Henri–Michaelis–Menten kinetics [59–61] used in enzymatic reaction models; or Monod kinetics [62] used in bioreactor models. These can be incorporated into an ODE surrogate model in their standard functional forms. If a process does not fall into one of these categories, there are several possible steps to identify an appropriate form of the corresponding ODE term, otherwise a generic form may be used as described in Case 2 below.

**Step 4**: _Approximate the control terms_. This is done by identifying a continuous representation of the given control problem and implementing it in the ODE surrogate model. This may be as straightforward as adding a standard control term '$Bu$' to the end of the right-hand side of the appropriate state-variable equation(s) or it might require adding control to a process formulation.

**Step 5:** _Parametrize the ODE model._ Before the ODE model can be used to find the solution for the ABM control problem, its parameters need to be estimated so that the ODE model approximates the ABM dynamics. This is done by optimizing the parameters of the ODE model against a collection of ABM simulation trajectories. Given that ABMs are often stochastic, several simulation runs for the same initialization should be averaged. The range of initializations should be chosen such that the entire domain of interest of the ABM is covered with trajectories[47]. Additionally, the ODE model should also be trained with trajectories for which control is exerted on each of the control variables, thus ensuring that the ODE model is a good approximation of the ABM in its uncontrolled state and under different levels of control. The level or magnitude of control used should be more than what is expected to be the optimal solution of the ABM control problem, which ensures that solving the control problem in the ODE model is an interpolation rather than an extrapolation problem. Initial guesses of the parameters may be obtained using the time course slope method [63,64], or model-specific parameter optimization methods, as is the case for Lotka–Volterra models [65]. We compute parameter estimates $\hat{p}$ using the least square method

$$\hat{p} = arg\underset{p}{min}(\|Y(u) - X(p,u)\|^2),$$ (Eq. 2)

where $p$ is the vector of all parameters of the ODE model, $u$ is the vector of all control parameters, $Y$ the vector of all timepoints over all trajectories of the ABM, and $X$ is the vector of the corresponding timepoints values obtained from the ODE model.

**Step 6:** _Solve the ABM control problem using the ODE approximation._ Standard methods from control theory can be used to solve the optimal control problem in the ODE approximation. This solution is then lifted back to the ABM. This is done by taking into consideration that now the solution may have to be discretized or the control inputs and outputs may require rescaling, depending on how these were approximated from the ABM into the ODE model (Step 4).

**CASE 2: Information is missing about the functional form of the model processes.**

**Input:** Information to determine the stoichiometric matrix of the ODE model, but not the functional form of the processes.

If sufficient information about the functional processes in the ABM is not available or the ABM processes are too complex, then either the functional forms will have to be reverse-engineered [66] or a generic functional form should be applied. Here we opted for the latter, which can be applied in a broad context. The only difference to Case 1 comes in Step 3, which we now describe.

**Step 3:** _Choosing the functional form of the processes._ If the functional form of one or more processes could not be identified in Case 1, then these will have to be identified by other means. Here, we present a biology-informed generic functional form, the power-law model.

Unlike in physics, it is often impractical to deduce concise functional expressions describing the evolution of biological processes from fundamental principles. In some cases, it is possible to derive semi-mechanistic representations of specific processes as is the case for the Henri–Michaelis–Menten approximation [59–61]. However, this seems to be more of an exception rather than a rule. Hence, when a functional representation is needed to describe a large range of different processes associated with a given ABM, there are only two other solutions: a generic function for which no guarantee exists of its correctness, or a canonical approximation.

There exist several canonical modeling frameworks in biology (as reviewed in [67]), including Lotka–Volterra systems [57,58], biochemical systems theory (BST) [68–72] and metabolic control analysis (MCA) [73,74]. Lotka–Volterra systems have limited usefulness in the context of general

ABMs since they lack the flexibility to capture non-linear processes. In MCA, lin-log functions are used to describe the vector field of an ODE system [75]. This modeling framework has been primarily employed for describing enzymatic reactions. In contrast to MCA, BST expanded from its initial area of applications that mainly involved biochemical systems to other biological fields, like biomedical applications [49,76–79]. In BST, the vector field of an ODE system is described by power laws. In BST, process $F_i$ is given by

$$F_j = \alpha_j \prod_{k=1}^{n} X_k^{g_{jk}},$$ (Eq. 3)

where $n$ is the number of state variables $X_k$ affecting the process $F_j$, $\alpha_j \in \mathbb{R}^+$ is the rate constant, and $g_{jk} \in \mathbb{R}$ are the kinetic orders. In BST, kinetic orders are real-valued parameters that express the dependence of $F_j$ on $X_k$. If a state variable has a positive effect on the process, the kinetic order is positive. State variables that inhibit a process have negative kinetic orders. Likewise, a state variable that does not affect a process will have a kinetic order of zero. If the dependency of the processes on state variables are not known, it is possible to just assume that all state variables of the system may regulate each of the processes of a system. Variable selection is then left to the optimization step (see Step 5 in Case 1). Otherwise, if some state variables are known not to be involved in certain processes, then the corresponding kinetic orders should be fixed to zero, thus reducing the number of parameters to be optimized.

## CASE 3. Control at a Steady State Without Information About ODEs.

Some ABMs exhibit specific regions in state space towards which trajectories (or averaged trajectories) tend to converge. This behavior is similar to stable steady states arising in ODEs. Unlike ODEs, however, stochastic ABMs usually do not converge towards a steady state. If an ABM control problem involves a steady state, and we aim to approximate the ABM in its vicinity, then we may use a linear approximation, essentially a first-order Taylor approximation:

$$\frac{dX}{dt} = J \cdot (X - X_0). \tag{Eq. 4}$$

When employing a steady-state approximation, it is important to verify that the ODE model is a good approximation of the ABM within the considered domain. Furthermore, the ODE approximation should be compatible with any control-induced displacements of the model trajectories. If the range of validity is too limited, then a second-order approximation may be used instead. That is,

$$\frac{dX_i}{dt} = \sum_{j=1}^{n} J_{ij}\left[X_j - (X_0)_j\right] + \frac{1}{2}(X - X_0)^T \cdot H_i \cdot (X - X_0), \tag{Eq. 5}$$

where $X$ is the vector of state variables, $X_0$ the steady state of the ABM, $J$ the Jacobian matrix, and $H_i$ the Hessian matrices of each differential equation $i=1,2,...,n$.

When choosing between low and high-order approximations, it is important to consider that the latter, while potentially more accurate, involve a greater number of parameters. A larger number of parameters can lead to increased complexity in generating the approximation and may require a larger dataset. While a first-order approximation will have $n^2$ parameters, a second-order approximation will have $\frac{1}{2}(3n^2+n^3)$ parameters, where $n$ is the number of state variables. Additionally, in second-order approximations, the corresponding ODE model may exhibit other steady states, and when control is exerted, the ODE system may be driven to or away from these regions. If the original ABM does not show evidence of any other steady states, then the solution is to select an ODE model during parameter estimation that does not have any other roots within the region of interest of the control problem.

**Step 1:** _Identify the state variables._ As in Cases 1 and 2, agent types or states have to be aggregated into state variables. However, since this approximation is not mechanistic, it may even be possible to represent only a subset of the agents as state variables and still achieve a good approximation.

Because there are no mechanistic functions being approximated, the primary factor determining how agents are mapped to state variables will be the underlying control problem. Therefore, any agent state that is distinct from all others, serving as an input to the control problem or having control exerted on it, must be represented by a separate state variable.

**Step 2:** _Determine the steady state_. Using several simulations of the ABM with different initializations and averaging the region of convergence of all trajectories will yield the steady state of the ABM ($X_0$).

**Step 3**: _Approximate the control terms_. This is done by identifying a continuous representation of a given ABM control problem and incorporating it in the steady-state ODE approximation.

**Step 4:** _Optimize ODE Parameterization for Control._ This step is equivalent to Step 5 in Case 1.

**Step 5:** _Control of ABM using an ODE approximation._ This step is equivalent to Step 6 in Case 1.

## CASE 4. S-System approximation in the absence of any information about the ODE structure.

We finally address cases in which it is unknown how state variables affect one another – examples include the stoichiometric matrix and the functional form of the processes that regulate interactions between state variables. As explained in the Introduction, we are primarily interested in solving control problems for biomedical applications, for which the S-system approximation [48,80] is a good candidate, given that it is mathematically tractable and consists of simple functions.

In an S-system representation [48,68–72,80,81], a system of ODEs is defined with each differential equation being equal to the difference of two power-law terms (Eq. 6). The first term, which is positive, aggregates all incoming processes, while the second term, which is negative, aggregates all outgoing processes. Thus, the evolution of state variable $X_i$ is described by

$$\dot{X}_i = \alpha_i \prod_{j=1}^{n} X_j^{g_{ij}} - \beta_i \prod_{j=1}^{n} X_j^{h_{ij}} \; for \, i = 1,2,\dots,n, \qquad \text{(Eq. 6)}$$

where, $n \in \mathbb{N}$ is the total number of state variables $X_i$, $\alpha_i$, $\beta_i \in \mathbb{R}^+$ are the rate constants and $g_{ij}$, $h_{ij}$ $\in \mathbb{R}$ are the kinetic orders.

The S-system representation (Eq. 6) models the effects that each state variable has on the positive and negative terms of the other state variables. The S-system representation does not require knowledge of the entire list of processes and interactions, as these are not explicitly modeled. Larger systems may be increasingly difficult to obtain, as the number of parameters in a system with $n$ state variables is $2(n+n^2)$.

**Step 1:** _Identify the state variables of the ODE_. This step closely mirrors step 1 of Case 3. Once the list of state variables has been determined, we can write down an ODE approximation in the form of an S-system (Eq. 6).

**Step 2**: _Approximate the control terms_. This step is equivalent to Step 3 of Case 3.

**Step 3:** _Optimize ODE Parameterization for control_. The procedure is the same as in Step 5 in Case 1. The rate constants ($\alpha$ and $\beta$) and kinetic orders ($g$ and $h$) are estimated using datasets generated from the ABMs and the parameters of the S-system model estimated using a least square estimation method (Eq. 2).

**Step 4:** _Control of ABM using an ODE approximation_. This step is equivalent to Step 6 in Case 1.

## Results

To illustrate the algorithm and test its performance, we considered two ABM control problems to examine how well each one of the described ODE surrogate modeling methods performs in identifying appropriate optimal control solutions. The models and the control problems were chosen for their relative simplicity, aiming to illustrate the main steps of the proposed surrogate modeling and control approach.

Since future medical digital twins are anticipated to be complex hybrid ABMs, the ABMs employed in this study serve as suitable foundational models for the development of analytical tools applicable to ABMs in general and medical digital twins in particular. Moreover, the optimization of treatment within a digital twin is conceptually similar to population-level control, rendering these models valuable test cases for the development of optimal control techniques tailored to ABMs.

The first ABM is based on the sheep-wolves-grass model [37], a generalization of two-species predator-prey models [22–24,82]. This, and related models, are commonly used in ecology and systems biology to describe the interactions and competitive dynamics among different species. Its mechanistic approximation aligns with the well-established Lotka–Volterra model. The second ABM we considered was designed to be more intricate, possessing features that pose challenges for both power-law models and the S-system approximations. This model represents a simplified metabolic network with five metabolites and four enzymes. The underlying processes are stochastic discrete representations of Michaelis–Menten dynamics. Due to the increased number of variables and the presence of non-linear processes, this model exhibited a higher level of complexity compared to the first ABM.

## Sheep-wolves-grass model

We considered the sheep-wolves-grass model as implemented in NetLogo [34], and increase the "world" size and the initial number of animals by a factor of 25, resulting in a grid of size 255×255, initially containing 1,250 wolves, and 2,500 sheep. Grass was always initialized as covering 50% of the world. Further implementation details are summarized in the Supplemental Information section. Below, we provide an overview of the datasets that we generated to train the ODE surrogate models. Different datasets are associated with different initial conditions, ABM parameters, and controls. In all ABM trajectories that we used to train the ODE surrogates, we averaged 100 realizations. For the simulations without control (datasets I and II), two different sets of initial conditions were used. For the simulations with control, we used the same initial condition as in dataset I and control either grass, sheep, or wolves (datasets III, IV and V).

The datasets without control were used to train the ODE models for the baseline dynamics of the ABM, whereas the datasets in which control was exerted on each species enable us to train ODE models that incorporate the ABM's behavior under control. The objective was to determine the constant rates at which wolves and sheep need to be removed to transition the system from its current steady state to a new one with 50% fewer wolves and 10% more sheep as compared to the original steady state. Additionally, we aimed to find a solution that minimizes the total number of animals removed. This resulted in a classical control problem with input matrix $B=diag(0, 1, 1)$ and control input $u=[0,-\kappa_2 Y,-\kappa_3 Z]^T$. Here, $X$ is the total amount of grass, $Y$ the total number of sheep, and $Z$ the total number of wolves. The quantities $\kappa_2$ and $\kappa_3$ are the removal rates that we wish to determine.

We employed five different ODE approximations to identify suitable control signals in the described ABM control problem:

1. A mechanistic approximation (Case 1), resulting in a Lotka–Volterra model.

2. A generalized mass action (GMA) model, where all seven processes were modeled using power laws involving all three variables (Case 2).

3. Linear and quadratic approximations at the steady state (Case 3).

4. An S-system model (Case 4), in which each differential equation was expressed as the difference between two power-law terms involving all three variables.

We parameterized each model against either datasets I and II, or datasets I-V, to study the effect of training the ODEs on datasets containing control information. Given that the ODE surrogates were to be evaluated for their ability to identify near-optimal control solutions, incorporating control information during the training stage transforms the control problem from an extrapolation into an interpolation task.

To validate the control solutions found by each ODE surrogate models, we employed a grid search to find the approximate mean optimal solution for the sheep-wolves-grass ABM control problem (black cross, Fig 3). Since the ABM never precisely reaches a steady state and instead, the three populations exhibit stochastic fluctuations around it, we also recorded all mean solutions located one standard deviation away from the target (orange dots, Fig 3). These suboptimal solutions illustrate the intrinsic level of noise present in the ABM and how it translates into the solutions of the control problem.

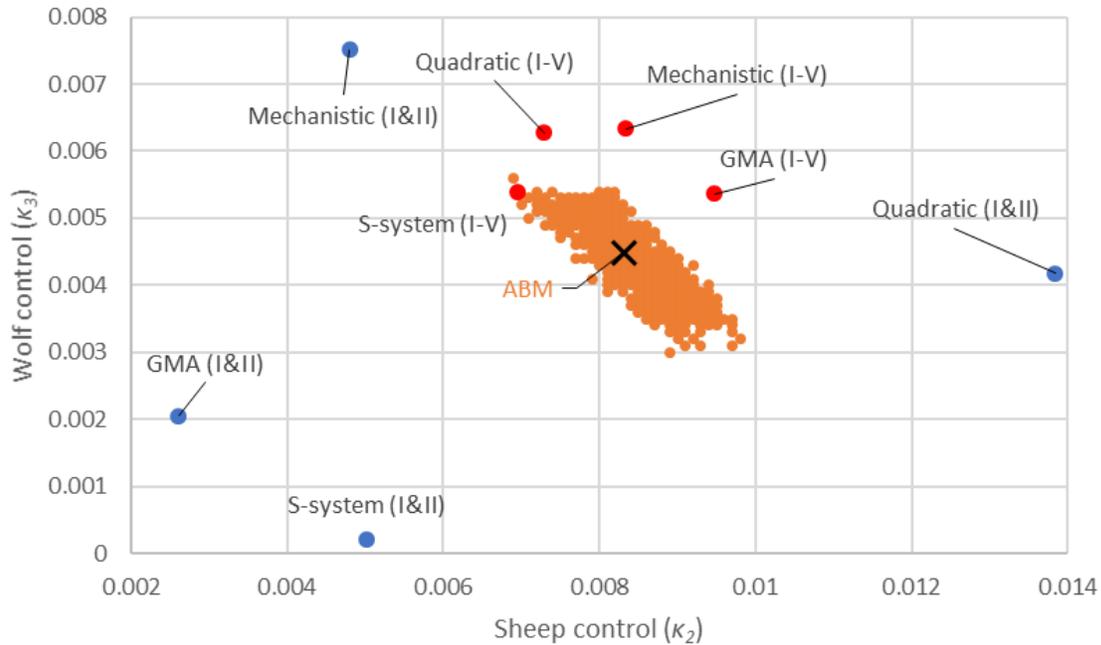

**Fig 3. Comparison of effectiveness of different ODE surrogate models for solving the sheep-wolves-grass ABM control problem.** The black cross marks the near-optimal solution ($\kappa_2 = 0.83\%$ and $\kappa_3 = 0.45\%$ per time step) for the sheep-wolves-grass ABM control problem as determined by a grid search (with a step of 0.0001 in both dimensions). Orange dots indicate suboptimal control solutions within one standard deviation from the target (a steady state with 50% fewer wolves and 10% more sheep compared to the original steady state). Blue and red dots show the control parameter values associated with the ODE surrogate models that have been calibrated against datasets I and II and datasets I-V, respectively. The best solutions were obtained for surrogate models parameterized with datasets containing control information (III-V). However, all four of these surrogate models (red dots) identified control solutions equally distant from the optimal one.

In Fig 3, we also show the control solutions as identified by each of the four ODE approximations, which we have parameterized either against datasets I and II (with no control information) or against datasets I-V (with control information). The linear approximation method (Case 3) delivered the worst performance in predicting the control solution and it was also the only approximation that was not able to simultaneously fit both datasets I and II. For this reason, the linear approximation was parameterized only against dataset I. All other approximations perform reasonably well at estimating the optimal solution of the ABM.

The primary factor influencing control performance was not the choice of approximations, but rather the datasets used to train the ODE models. Based on the data shown in Fig 3, we concluded that all four approximations, when parameterized against all five datasets (red dots), performed significantly better compared to when they were parameterized only against datasets I and II (blue dots). This highlights the critical importance of training the ODE surrogate models with simulations involving various levels of control to effectively solve ABM control problems. Training the ODE surrogate models on simulations where control was applied turns the estimation of the control solution from an extrapolation into an interpolation problem. We generated datasets IV and V by removing 2% of sheep and 1.5% of wolves per time step. These removal rates were set well above the approximate optimal solution of the ABM, which was identified as when 0.83% of sheep and 0.45% of wolves are concurrently removed per time step.

A possible explanation for the good performance of all four ODE surrogate models is that the mechanistic ODE model for the sheep-wolves-grass ABM aligns with a Lotka–Volterra model. This model essentially consists of a set of homogeneous second-order polynomial functions, which form a mass–action model. The other three ODE surrogates either incorporate the mechanistic model as a special case or closely resemble it. For instance, the quadratic approximation employs a set of second-order nonhomogeneous polynomial functions. The GMA approximation is constructed from a sum of power-law terms, of which mass-action serves as a special case (except for the grass growth term, which is $k_1 \cdot X - k_2 \cdot X^2$ in the mechanistic approximation and $\alpha_1 \cdot X^{g11} \cdot Y^{g12} \cdot Z^{g13}$ in the GMA approximation). The S-system is similar to the GMA, but is represented as the difference between two power-law terms, whereas the GMA model features three power-law terms in the differential equation describing the evolution of sheep.

## Metabolic pathway model

To further elucidate the differences between the different ODE surrogate models and to test their limitations, we next considered a second ABM, a metabolic pathway model for which the macroscopic mechanistic surrogate model would be given by an ODE system with Michaelis–Menten processes. In the described metabolic pathway model, we expected the ODE approximations outlined in Cases 2-4 to face difficulties in accurately representing the underlying dynamics. For example, ODE models based on power laws cannot capture saturation effects, which are characteristic of Michaelis–Menten processes. Additionally, S-system models are not good at modeling divergent and convergent pathways. The complexity of the considered metabolic dynamics could also pose difficulties for accurate approximation using first and second-order polynomials.

The metabolic pathway ABM included four reactions associated with five metabolites, and all reactions between enzymes, metabolites, and respective complexes were modeled at the elementary level (microscale). There were two agent types (metabolites and enzymatic complexes), five metabolites, four enzymes, and 12 enzyme-metabolite complexes. Metabolites moved ten times faster than enzymes or complexes, and whenever a metabolite is close to an enzyme or complex to which it may be bound, there was a probability that it may bind. Complexes could, at any time point, dissociate into their components. Enzymes formed complexes with their respective substrates, products, and regulators. Finally, enzymes complexed with their respective substrate could undergo catalysis and become a complex between the enzyme and the product. All four enzymatic reactions were modeled as irreversible.

Two collections of datasets were generated from this ABM. The first collection comprised two types of simulations: dataset I involved a single simulation where most of the pathway substrate was depleted, leading to the accumulation of two products (pathway operating in batch mode); and dataset II involved a single simulation where substrate was continuously supplied at a rate of one molecule per time step, and all metabolites were removed at a rate of 0.05% per time step (pathway operating in continuous mode as if in a continuous stirred tank reactor).

The second collection consists of three datasets, each resulting from the average of 100 simulations of the ABM under the same initial conditions. Datasets III and IV are based on averaging 100 simulations using the same parameters as datasets I and II, respectively. Dataset V is obtained by averaging 100 simulations under continuous mode with substrate being continuously supplied at a rate of 0.2 molecules per time step, while all metabolites were removed at a rate of 0.05% per time-step. In all continuous mode simulations, constant vessel volume was assumed, meaning that inflow and outflow match in volume, with only metabolites exiting while enzymes and enzymatic complexes remain within the vessel.

The control problem to be solved in the metabolic pathway model was the inference of the optimal inflow of substrate $S$ that minimizes the loss of the substrate $S$ and maximizes the amount of the products $R$ and $T$ at the outflow of the reactor. Mathematically, our goal was to identify the constant $Q_{in}$ that minimizes the loss function,

$$Loss(Q\ ) = \sum_{k=1}^{N_t} \frac{S_k}{R_k + T_k}, \quad \text{(Eq. 7)}$$

during a simulation run of $N_t$=50,000 time-steps, where $S_k$ is the concentration of the supplied substrate, at time step $k$, and $R_k$ and $T_k$ are the corresponding concentrations of the end-products of the pathway.

To compare the ability of all proposed ODE surrogates (mechanistic, GMA, S-system, quadratic and linear) to learn effective control solutions, we optimized them against the two collections of datasets ('I' will denote models optimized against datasets I and II, generated with a single simulation of each condition; and 'C' will denote models optimized against datasets III-V, generated by averaging 100 simulations of each condition).

To evaluate the predictions of each of the ODE approximations, the optimal inflow point was estimated for the ABM by performing a grid search between 0 and 1.0 with a step of 0.1. At each step, 100 simulation runs of the ABM were performed and averaged. The best inflow of substrate was found to be 0.7 (red square, Fig 4). The red line shows the mean loss function value at each of the tested inflow points of the ABM, and the orange band highlight where 75% of the simulations reside for each inflow point.

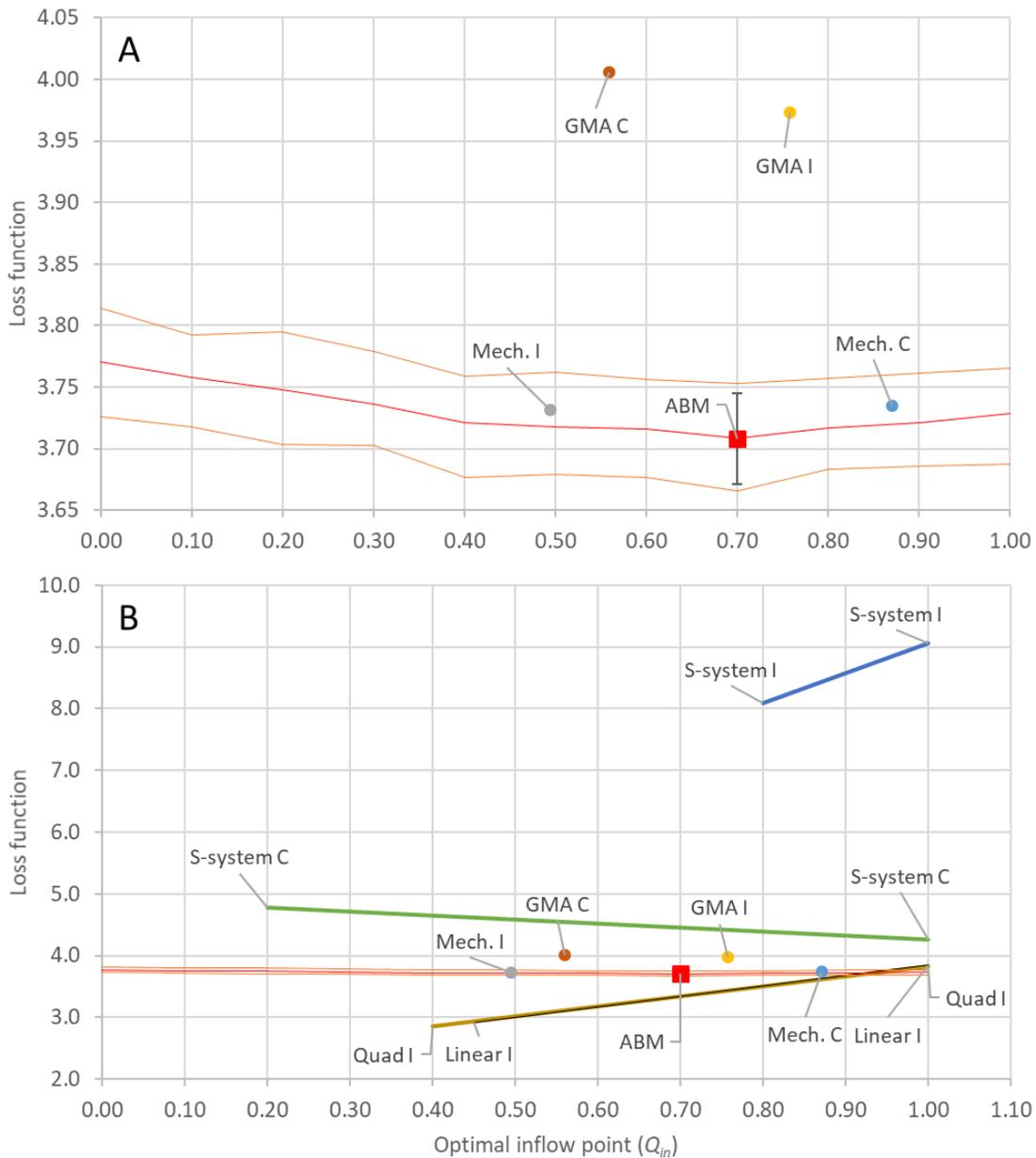

**Fig 4. Comparison of effectiveness of different ODE surrogate models for solving the metabolic pathway ABM control problem.** The red square shows the optimal inflow point and the corresponding mean loss function value as determined for the ABM by a grid search between 0 and 1.0 with a step size of 0.1, where in each step 100 simulations runs were averaged. The red line highlights the mean of each of the 100 simulation runs of the ABM and the orange line the 75% confidence band. Circles denote the predicted optimal inflow and corresponding loss function value for each ODE surrogate. ODE models that did not exhibit a minimum within the 0 to 1.0 domain have their domain of integrability shown with a line. The line depicts the range of loss function values predicted by the approximation. The S-system I performed worst, as it could only be integrated between 0.8 and 1.0, and in that range predicted loss function values between 8 and 9. While S-system C, Quad I, and Linear I, all resulted in models with a larger domain over which they could be integrated, neither had a minimum within their respective domains. GMA I was the ODE surrogate that best predicted an optimal inflow of substrate closest to the ABM and Mech. I best predicted the loss function value of the ABM at the optimal inflow point.

Based on the findings presented in Fig 4, we drew the following conclusions: the GMA approximation, when parameterized against datasets I and II, performed the best in predicting the optimal substrate inflow. On the other hand, the mechanistic approximation, also parameterized against datasets I and II, performed well in estimating the loss function value. Neither of the S-system approximations displayed a minimum within the control parameter range of [0.2, 1]. Additionally, the S-systems generated against datasets I and II could only be solved between 0.8 and 1.0 due to stiffness issues. Similarly, both the quadratic and linear approximations were unable to predict an optimal inflow, as they could not be integrated across the entire domain and did not exhibit a minimum within the region where they could be integrated. However, the S-system parameterized against the averaged simulations (S-system C, derived from datasets III-V) demonstrated a notably better estimate of the loss function compared to the S-system obtained for datasets I and II (S-system I). This suggests that the S-system is more susceptible to the noise present in datasets I and II, which is reduced in the datasets resulting from averaging 100 simulations of the ABM (datasets III-V).

## Discussion

Many problems in medicine are control problems. But directly testing various control approaches, such as different treatment protocols, in the laboratory or in clinical studies is often impractical. To address this, medical digital twins offer a solution for developing effective treatments based on *in silico* treatment optimization [83–86]. The mathematical models underlying them can be diverse. One commonly used model type is ABMs, used in biomedicine to simulate complex interaction among multispecies populations. Controlling (or optimizing) ABMs is challenging, however, due to their underlying stochastic and high-dimensional dynamics and rule-based rather than equation-based structure. While traditional control methods are well-developed for ODE systems, they are not directly applicable to ABMs. To bridge the gap between standard control-theoretic approaches and ABM simulation-based methods, we proposed different types of ODE surrogate models to approximate the dynamics of a given ABM control problem. In addition to their use in control scenarios, they can also serve forecasting purposes by modeling their uncontrolled evolution. The proposed methods allow us to identify effective controls in the ODE surrogate and apply them back to an ABM. We consider this work as a small first step toward solving the general problem of surrogate modeling and optimization for medical digital twins. In summary, we developed five different approaches capable of generating ODE surrogate models from ABMs with dynamical and steady-state regimes. We employed all approximation methods on two typical ABM systems featuring Lotka–Volterra and Michaelis–Menten dynamics, which are commonly encountered in a broad spectrum of applications.

For approximations under a steady-state regime, we present two approaches: the linear and the quadratic approximations. These approaches do not require any previous knowledge of the underlying mechanisms of a given ABM control problem and they may allow one to construct approximations using only a small subset of all ABM variables. Although the linear approximation does have advantages in terms of optimal control theory, since optimal control problems associated with linear dynamical systems have analytical solutions [6–8], it was not a good approximation in either of the examples presented. A natural improvement is to employ a second-order approximation method, which we refer to as the quadratic approximation. This approximation performed better in the sheep-wolves-grass control problem than the linear approximation, but increasing the order led to the introduction of unstable steady states. Hence, it became difficult to optimize the control parameters. As a solution to this problem, we identified the best quadratic approximation that fit the ABM data and contained no other steady states within the domain of the control problem. With this additional optimization constraint, the quadratic approximation performed as well as the remaining approximations in the sheep-wolves-grass ABM.

In S-system models, each state variable is characterized by two power-law terms, one for the "inflow" and one for the "outflow" [48,80,81]. Additionally, all state variables of the system can be included in both of these terms. Given that power-law terms are linear representations of the processes in log-log space, this non-linear behavior may give it an advantage over a linear approximation. On the other hand, an S-system model will have $2(n+n^2)$ parameters in an $n$-dimensional ODE system, whereas a linear approximation will only have $n^2$ parameters. Thus, the extra non-linearity comes at the expense of more parameters. The quadratic approximation has $\frac{1}{2}(3n^2+n^3)$ parameters, which is even more than in the S-system. This makes the S-system a compromise between the linear and the quadratic approximations in terms of parameter numbers. All of these representations may gain from being parameterized with Lasso and related regularization approaches [87], which can help identify the most parsimonious model that fits the available data.

Within the mechanistic class of approaches, we considered two surrogate modeling methods: the mechanistic approximation and the GMA approximation. Among all of the ODE surrogate model methods proposed in this work, the mechanistic approach bears the closest resemblance to the equation learning approach (EQL) studied by Nardini *et al.* [88]. One key difference between the mechanistic method used in our work and the EQL method lies in how Nardini *et al.* construct a function library based on process representations inferred from a given ABM, with differential equations expressed as a linear combination of these library terms. In contrast, our approach involves utilizing an inferred stoichiometric matrix to determine which process representations have to be considered in each differential equation. In cases such as epidemiological and ecological

ABMs, where both approaches are likely to result in ODEs approximated by a linear combination of polynomial terms, the resulting ODEs may seem quite similar. However, our method distinguishes itself by employing the stoichiometric matrix to determine term selection, leaving the optimization process exclusively for the refinement of process parameters. In the EQL approach, term selection and process parameters are left to the optimization process.

The GMA method represents a different paradigm. In this approach, rather than inferring the appropriate functional representation of the processes mechanistically, the functional forms are approximated by power laws and assumed to be dependent on all state variables. During the parameterization phase, a regularization method like Lasso may be used to identify a parsimonious model that is still able to capture the evolution of ABM-generated data. In GMA models, we identify not only which state variables each process depends on, but also the sign (positive for activation and negative for inhibition) and strength (the absolute value) of each dependency.

All of the approaches presented here depend on the optimization of an ODE model to ABM-generated data. Identifying an appropriate ODE surrogate model might not always be feasible. This can occur due to either the inherent limitations of the chosen surrogate model or difficulties in finding a suitable parameter set through optimization. If one expects limitations that are due to a specific optimizer choice, exploring alternative optimization techniques may provide a solution. Similar to related problems within machine learning [89], there exists a spectrum of algorithms for parameter optimization, ranging from local and usually deterministic techniques to global and stochastic ones. Furthermore, there are problem-specific approaches that require the user to understand the nature of the optimization problem and have knowledge about the most appropriate optimizers [67,80,90].

In the first example involving the sheep-wolves-grass ABM, we examined whether using training data with varying levels of control improved the predictive power of the ODE approximations. Our findings showed that optimizing ODE approximations using simulations with different control levels significantly enhanced their ability to identify suitable control signals. By training ODE approximations on ABM datasets obtained with both higher and lower levels of control relative to the true optimum, we effectively transform the problem of inferring the optimal control from an extrapolation to an interpolation task. This transformation is contingent on knowing or being able to assume or infer the domain of the control problem.

In the metabolic pathway ABM, we compared the effectiveness of averaged ABM simulations with single simulations. The results provided a more detailed picture of the performance differences between the various ODE surrogate models. For the mechanistic and GMA surrogates that provided relatively accurate estimates of the optimal control input, averaging simulations did not confer any notable advantage. In the case of the S-system surrogate, it did not perform well in estimating a suitable control signal, but it did demonstrate an improved ability to estimate the loss function when we employed averaged training data.

Among all the methods studied in this work, the GMA appears to be the most practical choice. While the mechanistic approximation is likely to yield the most accurate surrogate model, it is impractical to generate mechanistic surrogates for complex multiscale ABMs like the ones considered here and in related works (see, *e.g.*, [9,20,21]). The GMA approach bypasses the need for detailed mechanistic information about the ABM, focusing solely on understanding the structure of the processes. This is a less complex task compared to capturing process representations. In the examples provided here, the GMA approach performed as effectively as the mechanistic approach (see Table 1 for a summary of the properties of the different methods).

On the other end of the spectrum lies the S-systems approach, which does not require detailed mechanistic information about a given ABM (Table 1). Instead, it requires knowledge of which agent types or states will be aggregated into each of the state variables of the surrogate model. However, in the metabolic pathway example, intentionally designed to challenge the S-system approach due to its complex structure and functional representation, the S-system did not perform well. None of the datasets used were able to generate an S-system surrogate capable of predicting the optimal control of the ABM. These findings underscore the importance of exploring approaches that fall between fully mechanistic and non-mechanistic surrogate models. Hybrid approximations may prove to be especially valuable in such cases.

**Table 1. Overview of advantages and disadvantages of the different ODE metamodels.**

| Surrogate Model | Advantages | Disadvantages |
|---|---|---|
| **Mechanistic** | • Most accurate approximations.<br>• Accurate over a wide range of the state space. | • Hard to formulate for complex ABMs.<br>• Requires detailed mechanistic knowledge of the inner workings of the ABM.<br>• ODEs may have a complex mathematical structure. |
| **GMA** | • No need for process-specific approximations; each process is formulated by a power law involving system state variables.<br>• Determining the stoichiometric matrix is simpler than mechanistically approximating the processes.<br>• Determining ODEs becomes straightforward once the stoichiometric matrix is known.<br>• Can be semi-mechanistic by employing mechanistic approximations for some processes and representing others with power laws.<br>• Small to moderate number of parameters. | • Still requires inferring the stoichiometric matrix of the ABM.<br>• Power laws face challenges when approximating processes that reach saturation because power laws with positive kinetic orders tend to approach infinity. |
| **Taylor expansion at the steady state** | • With an accurate linear approximation, one can apply established control-theoretic methods for linear systems.<br>• Straightforward to determine the ODEs. | • Unlikely to be a good approximation at lower orders and may only be accurate in specific regions of the state space.<br>• Large number of parameters at high orders. |
| **S-system** | • Straightforward to determine the ODEs with a rule-based (canonical) approach.<br>• Requires no mechanistic understanding of the ABM.<br>• Moderate number of parameters.<br>• Homogeneous ODEs with favorable mathematical properties, making steady-state solutions easy to obtain. | • Power laws face challenges when approximating processes that reach saturation because power laws with positive kinetic orders tend to approach infinity.<br>• May encounter challenges when approximating systems with multiple processes affecting the same agent state. |

In conclusion, we introduced four distinct families of approximation methods that can be employed to solve ABM control problems. We focused on ABMs with fully dynamic transients and those with stable steady states. The mechanistic approaches proposed here offer a key advantage in utilizing a stoichiometric matrix as a foundational structure for representing interactions and events within a given ABM. One major advantage of using mechanistic approximation methods is their interpretability. However, we also obtained promising results using different levels of generic ODEs. Among the approaches examined, the GMA and mechanistic methods were the most effective ones in identifying suitable control signals for a given ABM control problem.

There are several promising directions for future research. One avenue involves assessing the effectiveness of the approximation methods developed in this study on additional ABMs, particularly in the context of digital twin models integrated with patient data [20,21,83–85]. Furthermore, future investigations may focus on the application of data assimilation techniques to dynamically refine existing surrogate models as new data become available [91,92], especially in cases where ABMs are being reparametrized to account for updated patient information. Finally, it would be valuable to explore alternative ODE approximators or optimization methods focused on directly controlling specific aspects of an ABM [93–97]. These research paths will not only contribute to enhancing our ability to control ABMs but may also help develop more effective control approaches in biomedicine and related fields in general[98].

## Acknowledgments

The authors are indebted to Dr. Anna Niarakis and the organizing committee for organizing a three-week workshop on "Building Immune Digital Twins" where the authors were able to work in person on this project. This workshop was supported by the Institut Pascal, University of Paris-Saclay, France via the program Investissements d'avenir, ANR-11-IDEX-0003-0, and by Genopole.

## Funding

The funders had no role in study design, data collection and analysis, decision to publish, or preparation of the manuscript.

National Institute of Health grant R01 GM127909 (RL, LLF)

National Institute of Health grant R01 AI135128 (RL, BM, LLF)

Defense Advanced Research Projects Agency grant HR00112220038 (RL, LLF)

Army Research Office grant W911NF-23-1-0129 (LB)

hessian.AI (LB)

National Institute of Health grant R01 HL169974 (RL, BM)

## Author contributions

Conceptualization: RL, LLF, LB

Methodology: RL, LLF, LB

Investigation: RL, LLF, LB

Visualization: LLF, LB

Writing—original draft: LLF

Writing—review & editing: RL, LLF, LB, BM

**Competing interests:** All authors declare they have no competing interests.

**Data and materials availability:** The code and datasets used in this work are available on GitHub: https://github.com/LaboratoryForSystemsMedicine/Metamodeling-and-Control-of-Medical-Digital-Twins_2024

**Supplementary Text 1**

## <u>The sheep-wolves-grass model</u>

The process of approximating an ODE to an ABM will be illustrated using the sheep-wolves-grass version of the Wolf Sheep Predation model[1] supplied with NetLogo[2]. The model was used as in the NetLogo database, with the exception that the size of the world was increased 25-fold, from a size of 51×51 to 255×255. Similarly, the initial populations were also increased 25-fold (sheep were increased from 100 to 2,500, and wolves from 50 to 1250), thus maintaining the initial density of agents in the world (See file 'Wolf Sheep Predation bigworld.nlogo'). Every patch in the world has a 50/50 probability of being available as green grass at the start of a simulation per definition of the original model, and this parameter was not changed. This bigger world made the simulations less noisy and allowed them to be more robust against extinction. Averaging was done by accumulating 100 runs of the same instantiation, which resulted in a similar amount of noise as with 2,500 simulations of the original 51×51 model. Under these conditions, this model has a stable steady state that is reached in approximately 300 time-steps, after which the total grass will remain around 24,290, total number of sheep around 4,001, and total number of wolves around 1,917.

To examine all ODE surrogate models, including those designed for a steady state and those not requiring steady state conditions, we formulated a control problem that targets a steady state. The control problem used is the following: We want to determine the number of wolves and sheep that need to be removed in order to move the steady state to a new point where there are only 50% of the wolves present and sheep go up to 110% relative to the original steady state. Additionally, we want the solution that results in the removal of the smallest number of animals. This results in a classical control problem with input matrix $B=diag(0,1,1)$, and control input $u=[0,-\kappa_2 Y,-\kappa_3 Z]^T$, where $X$ is the total amount of grass, $Y$ the total number of sheep and $Z$ the total number of wolves.

Dataset I (Fig S1, w/o control; dataset created by simulation of the model in the file 'Wolf Sheep Predation bigworld.nlogo') was generated using 1,250 initial wolves, 2,500 initial sheep, and the original distribution of grass. Dataset II differs from dataset I by having initially more sheep and wolves but less grass. Dataset II (dataset created by simulation of the model in the file 'Wolf Sheep Predation bigworld 2nd DataSet.nlogo') was generated by shifting the original model parameters to different values (sheep energy gain from food from 4 to 5, sheep reproduction rate from 4% to 5% per timestep, and wolf reproduction rate from 5% to 1% per timestep). After 1,000 timesteps when the system reached a new steady state, the parameters were changed back to the original values (4, 4, and 5, respectively) and dataset II recorded until the system settled in its original steady state (i.e., the same as in dataset I). In dataset II, all three species exhibit wider fluctuations than in dataset I (Fig S1).

The datasets with control (III-V, Fig S1) were generated using the same initial condition and parameters as for dataset I. The simulations were conducted over 1,000 timesteps to ensure that the ABM reached a steady state. We then exerted constant permanent control on each of the species. In dataset III (dataset created by simulation of the model in the file 'Wolf Sheep Predation bigworld_ConGrass2.nlogo'), grass control involved removing 2% of the available grass per timestep. For dataset IV (dataset created by simulation of the model in the file 'Wolf Sheep Predation bigworld_ConSheep2.nlogo'), sheep control was associated with a 2% removal of sheep per timestep. In the case of wolves (dataset V), control involved a 1.5% removal per time-step, as a 2% removal led to their extinction (dataset created by simulation of the model in the file 'Wolf Sheep Predation bigworld_ConWolves1.5.nlogo'). Subsequently, we ran the simulations for datasets III, IV, and V for an additional 500 timesteps while applying the described controls. This ultimately led to the dynamics reaching new steady states (Fig S1).

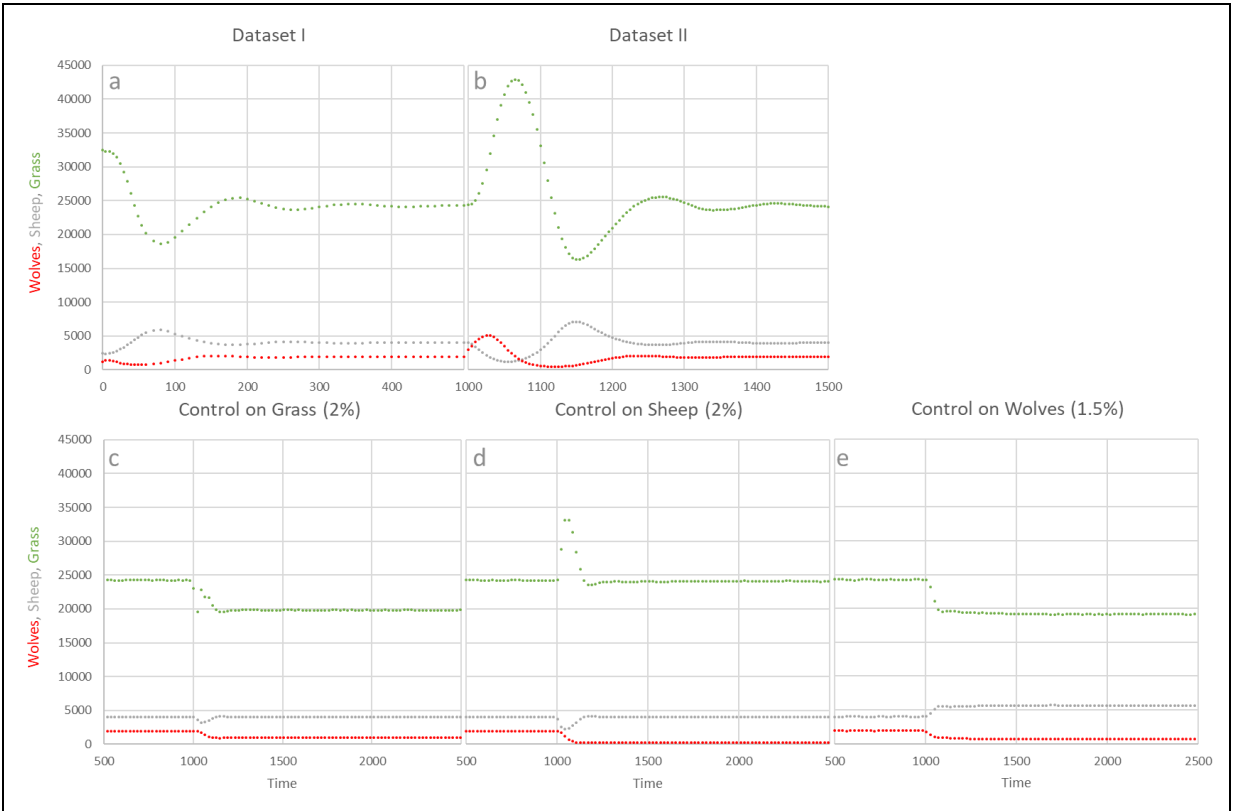

**Fig S1. Training datasets for all surrogate models of the sheep-wolves-grass ABM.** Datasets I and II (panels a and b) were generated with two different initial conditions. Panels c, d, and e were generated starting from the same initial conditions as dataset I (panel a) simulating until timestep 1,000, and then either 2% of grass was removed, 2% of sheep, or 1.5% of wolves.

## Case 1 – Mechanistic approximation

**Steps 1, 2, and 3**: Analysis of the ABM shows that each patch of grass takes 30 timesteps to regrow after being consumed and the world has a maximum size, carrying capacity for grass, of 65,025. Sheep gain energy from grass consumption at a fixed rate, have a fixed probability of breeding with energy being equally divided between offspring and parent, and lose energy at a fixed rate which results in death when energy is depleted. Wolves gain energy from sheep consumption at a fixed rate, have a fixed probability of breeding with energy being equally divided between offspring and parent, and lose energy at a fixed rate which results in death when energy is depleted. All interaction processes depend only on the co-localization of the two agents and are independent of all agent attributes including position. However, sheep and wolves do have a death process that is dependent on their energy, which could require sheep's and wolves' energies to be represented by state variables (Fig S2). Because energy acts as an inhibitor for the death of sheep and wolves, and inhibitory processes are not well-represented by mass action kinetics, we simplified the model by reassigning the negative impact of food (grass for sheep and sheep for wolves) on their mortality to a positive effect on their respective growth. This adjustment allows us to not include energy terms for both sheep and wolves in our ODE model (Fig S2). The resulting ODE model is

$$\frac{d}{dt}\begin{bmatrix} X \\ Y \\ Z \end{bmatrix} = M \cdot F, \qquad M = \begin{bmatrix} 1 & -1 & 0 & 0 & 0 & 0 & 0 \\ 0 & 0 & 1 & -1 & -1 & 0 & 0 \\ 0 & 0 & 0 & 0 & 0 & 1 & -1 \end{bmatrix},$$

$$F = \begin{bmatrix} k_1 \cdot X - k_2 \cdot X^2 \\ k_3 \cdot X \cdot Y \\ k_4 \cdot X \cdot Y \\ k_5 \cdot Y \\ k_6 \cdot Y \cdot Z \\ k_7 \cdot Y \cdot Z \\ k_8 \cdot Z \end{bmatrix},$$

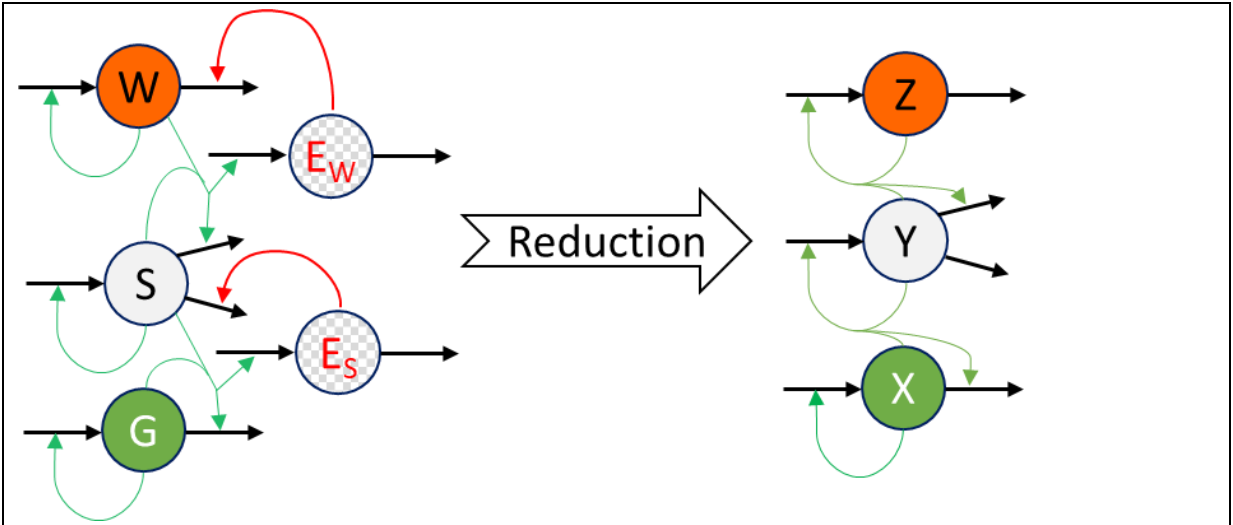

**Fig S2. Model reduction from the sheep-wolves-grass ABM to the ODE surrogate model.** In the ABM, the energy of sheep ($E_S$) and wolves ($E_W$) agents prevents (inhibits) their death. When the energy reaches zero, the agent dies. Generation of offspring is an event that occurs at every timestep and depends only of the probability of reproduction. After reproduction, energy is divided between parent and offspring. In order to approximate a mechanistic mass action ODE model, the inhibitory effect of energy of the agent on its death was reassigned to a positive effect on the growth of the population. W, S, and G denotes wolves, sheep and grass agents in the ABM. Z, Y, and X denote wolves, sheep and grass populations in the ODE model.

(Eq. 1.1)

or simply

$$\dot{X} = k_1 \cdot X - k_2 \cdot X^2 - k_3 \cdot X \cdot Y$$
$$\dot{Y} = k_4 \cdot X \cdot Y - k_5 \cdot Y - k_6 \cdot Y \cdot Z,$$
$$\dot{Z} = k_7 \cdot Y \cdot Z - k_8 \cdot Z$$

(Eq. 1.2)

where X, Y, and Z are the total amounts of grass, sheep, and wolves, respectively, in the world, and $k_i$ are the rate constants of the different processes.

**Step 4:** As described above, the control of the ABM is done by removal of a fixed percentage of the agents per timestep. This is easily approximated in an ODE by linear terms (Eq. 2).

$$\dot{X} = k_1 \cdot X - k_2 \cdot X^2 - k_3 \cdot X \cdot Y$$
$$\dot{Y} = k_4 \cdot X \cdot Y - k_5 \cdot Y - k_6 \cdot Y \cdot Z - \kappa_2 \cdot Y$$
$$\dot{Z} = k_7 \cdot Y \cdot Z - k_8 \cdot Z - \kappa_3 \cdot Z$$

(Eq. 2)

**Step 5:** To compare the effect of using datasets obtained with and without control, two parameterizations were obtained. The first parameterization was done against datasets I and II (Fig S3, (see file 'SWG_Case1_Mech.I_II.m')), and the second parameterization against all five datasets I-V (Fig S4, see file 'SWG_Case1_Mech.I_V.m'). Both parameterizations were done using least-square non-linear regression. An initial guess of the parameters was obtained using the time course slope method[3,4]. Alternatively, other methods specific for parameter optimization of Lotka–Volterra models may also be used[5]. The ODE models obtained are in good agreement with the corresponding datasets (Figs S3 and S4).

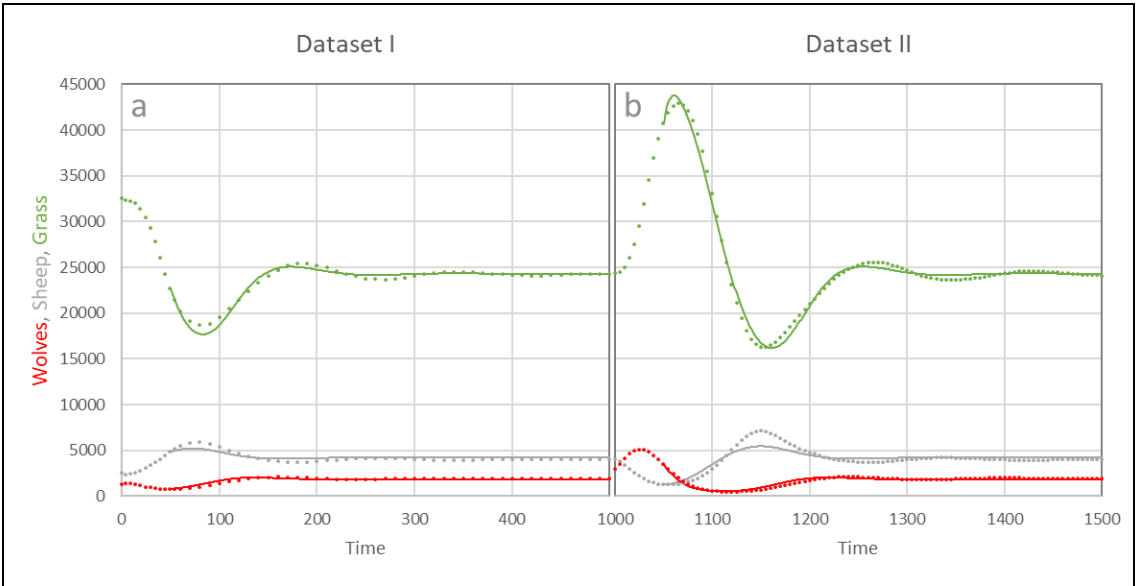

**Fig S3. Fit of the mechanistic approximation (Case 1) to the ABM datasets I&II.** The ODE model is shown as solid lines and training data as colored markers.

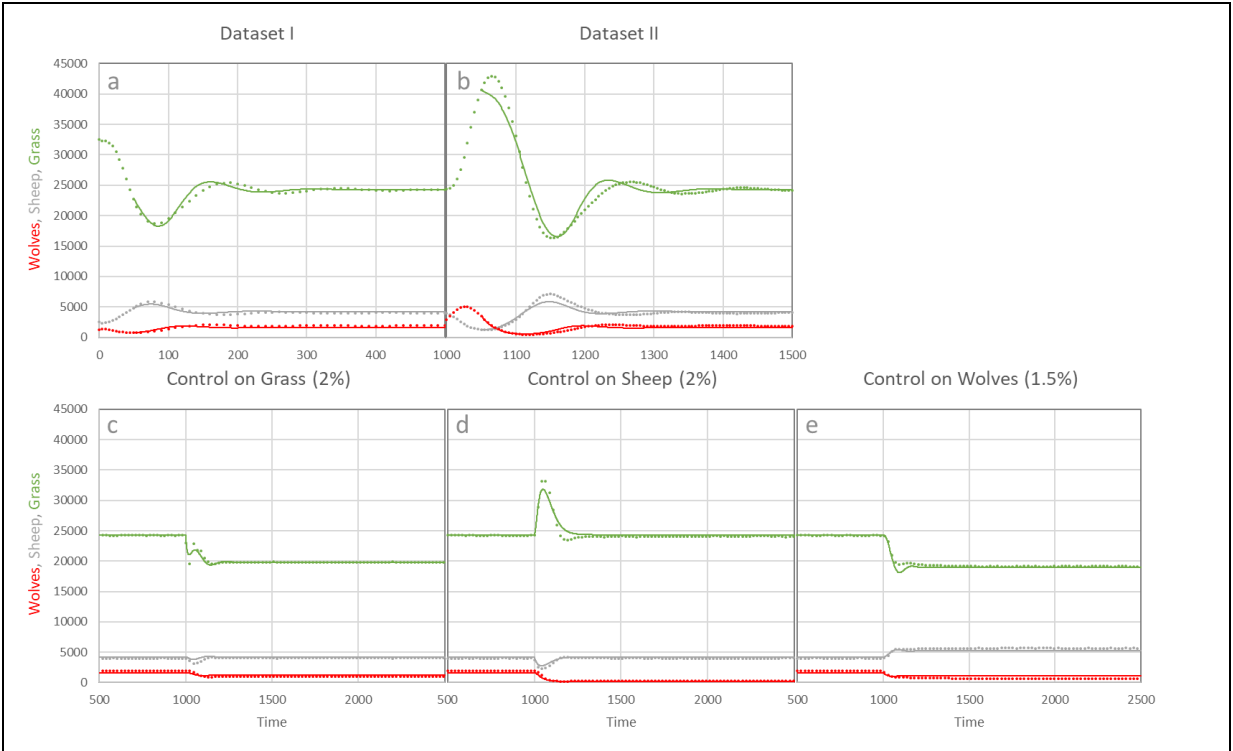

**Fig S4. Fit of the mechanistic approximation (Case 1) to the ABM datasets I-V.** The ODE model is shown as solid lines and training data as colored markers.

**Step 6:** Solving the control problem. Using the model above (Eq. 2) and the two parameterizations, $\kappa_2$ and $\kappa_3$ were determined so that this model (Eq. 2) has a steady state with 50% of the wolves and 110% of the sheep relative to the original model (Eq. 1.2). The results (mechanistic (I and II) and mechanistic (I-V)) are plotted in comparison with the best solutions found for the ABM, determined by performing a grid search on $\kappa_2$ and $\kappa_3$ (Fig S5).

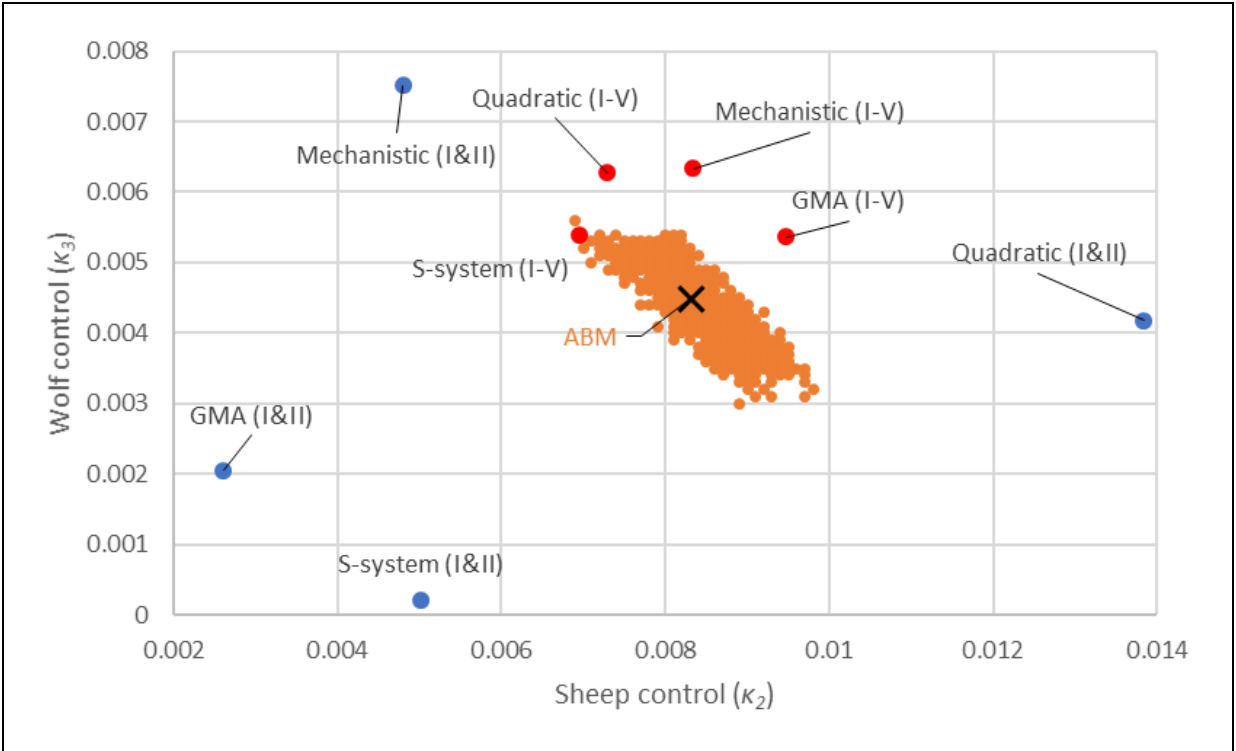

**Fig S5. Comparison of effectiveness of different ODE surrogate models for solving the sheep-wolves-grass ABM control problem.** The black cross marks the optimal solution ($\kappa_2$ = 0.83% and $\kappa_3$ = 0.45% per timestep) for the sheep-wolves-grass ABM control problem as determined by a grid search (with a step of 0.0001 in both dimensions). Orange dots indicate suboptimal control solutions within one standard deviation from the target (a steady state with 50% fewer wolves and 10% more sheep compared to the original steady state). Blue and red dots show the control parameter values associated with the ODE **surrogates** that have been calibrated against datasets I and II and datasets I-V, respectively. The best solutions were obtained for **surrogates** parameterized with datasets containing control information (III-V). However, all four ODE **surrogates** identified control solutions (red dots) equally distant from the optimal one.

**Case 2 – GMA approximation**

Depending on the complexity of the ABM being approximated, it might not be possible or desirable to perform mechanistic approximations to each and every process. One way to avoid having to assume or deduce mechanistic formulations is to use a canonical approach. In biochemical systems theory (BST) *(61,62,73)*, all processes are canonically represented by power laws

$$F_i = \alpha_i \prod_{j=1}^{m} X_j^{g_{ij}}. \qquad \text{(Eq. 3)}$$

In BST, if each process of a system is approximated as a power law this is referred to as a generalized mass action (GMA) model. On the other hand, if all processes into each state variable are approximated to a power law and all processes out of each state variable are approximated to another power law then this is referred to as an S-system model.

**Step 3:** The difference relative to case 1 will be that here the processes are represented by power laws of all three state variables. Thus, no assumptions are being made on which variables each of the processes depends on, although this could be done and would result in fewer parameters. Rather this dependence will be inferred from the datasets by optimization. The GMA ODE is given by

$$\frac{d}{dt}\begin{bmatrix} X \\ Y \\ Z \end{bmatrix} = M \cdot F, \qquad M = \begin{bmatrix} 1 & -1 & 0 & 0 & 0 & 0 & 0 \\ 0 & 0 & 1 & -1 & -1 & 0 & 0 \\ 0 & 0 & 0 & 0 & 0 & 1 & -1 \end{bmatrix},$$
$$F_i = \alpha_i \cdot X^{g_{i1}} \cdot Y^{g_{i2}} \cdot Z^{g_{i3}}, \qquad i \in \{1,2,3,4,5,6,7\},$$

(Eq. 4)

which shares the same stoichiometric matrix (M) with case 1 (Eq. 1.2).

**Step 4:** Control terms were approximated as in case 1, which yields a similar model

$$\frac{d}{dt}\begin{bmatrix} X \\ Y \\ Z \end{bmatrix} = M \cdot F - \begin{bmatrix} 0 \\ \kappa_2 \cdot Y \\ \kappa_3 \cdot Z \end{bmatrix},$$

$$M = \begin{bmatrix} 1 & -1 & 0 & 0 & 0 & 0 & 0 \\ 0 & 0 & 1 & -1 & -1 & 0 & 0 \\ 0 & 0 & 0 & 0 & 0 & 1 & -1 \end{bmatrix}, \quad F_i = \alpha_i \cdot X^{g_{i1}} \cdot Y^{g_{i2}} \cdot Z^{g_{i3}},$$

(Eq. 5)

where $i \in \{1, 2, 3, 4, 5, 6, 7\}$.

**Step 5:** As in the mechanistic approximation (case 1), we optimized the model against two sets of datasets for comparison. In the first (Case 2.1, GMA (I and II)), since not all variables are expected to regulate each of the processes, an L1-regularization term was added to the objective function that leads to the selection of the best fitted model with the fewest number of kinetic orders different from zero. This L1 approach is similar to what is used in the LASSO method[6]. We performed this optimization using only datasets I and II. In the second approach (Case 2.2, GMA (I-V)), we took advantage of all parameters and performed the optimization (regular non-linear optimization) against all five datasets (I-V). Datasets III-V were used by exerting control on each of the variables by setting *($\kappa_1$, $\kappa_2$, $\kappa_3$)* to either [0.02, 0, 0], [0, 0.02, 0], or [0, 0, 0.015], respectively, in equation Eq. 6.

$$\frac{d}{dt}\begin{bmatrix} X \\ Y \\ Z \end{bmatrix} = M \cdot F - \begin{bmatrix} \kappa_1 \cdot X \\ \kappa_2 \cdot Y \\ \kappa_3 \cdot Z \end{bmatrix},$$

$$M = \begin{bmatrix} 1 & -1 & 0 & 0 & 0 & 0 & 0 \\ 0 & 0 & 1 & -1 & -1 & 0 & 0 \\ 0 & 0 & 0 & 0 & 0 & 1 & -1 \end{bmatrix}, \quad F_i = \alpha_i \cdot X^{g_{i1}} \cdot Y^{g_{i2}} \cdot Z^{g_{i3}},$$

(Eq. 6)

where $i \in \{1, 2, 3, 4, 5, 6, 7\}$.

The first GMA model (Case 2.1, see file 'SWG_Case2_GMA.I_II.m') fitted well both datasets I and II (Fig S6). Interestingly, three of the kinetic orders were not needed for model fit, and so in this model, grass does not regulate $F_5$, sheep removal by wolves, and $F_7$, death of wolves. This latter process is also not dependent on sheep (Table S1). This effectively lowered the number of parameters to 25. Three kinetic orders were found to be negative: (i) grass' effect on grass growth, (ii) grass' effect on death of sheep, and (iii) wolves' effect on grass consumption by sheep (Table S1). These are interesting results as grass does technically inhibit the death of sheep, since in the ABM lack of energy causes sheep to die, and results confirm that the ABM simulations contained evidences of grass inhibiting the death of sheep. The inhibition of sheep consumption of grass by wolves ($g_{23}$ in $F_2$) makes sense only as an indirect effect, since wolves do not play an explicit role in that process, but higher levels of wolves do lead to decreased levels of sheep, which then eat less grass. Interestingly, the process of wolf removal was not found to be inhibited by sheep, which was expected given that in the considered ABM wolves die only due to a lack of energy.

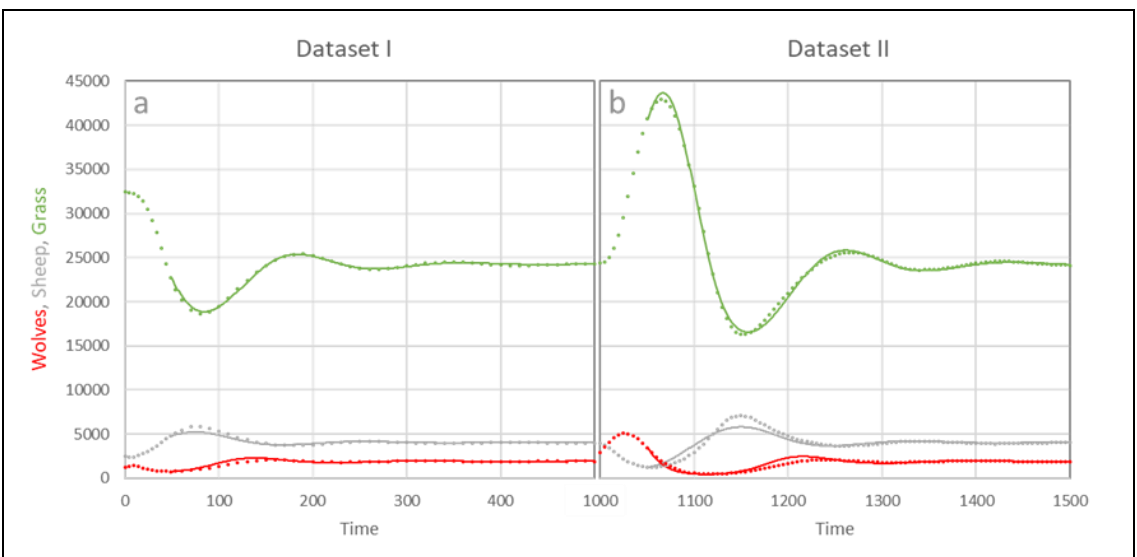

**Fig S6. Fit of the GMA approximation (Case 2) to the ABM datasets I&II.** GMA approximation (solid lines) and corresponding training data (colored markers).

Table S1. Parameters obtained for the processes of the canonical ODE model. All values were rounded to two decimal places.

| Processes | Rate constants | Kinetic Orders | | |
|---|---|---|---|---|
| ($i$) | ($i$) | $g_{i1}$ | $g_{i2}$ | $g_{i3}$ |
| $P_1$ | $7.08 \cdot 10^2$ | -0.13 | 0.07 | 0.14 |
| $P_2$ | $2.74 \cdot 10^{-5}$ | 1.01 | 0.92 | -0.06 |
| $P_3$ | $1.18 \cdot 10^{-6}$ | 0.94 | 0.99 | 0.06 |
| $P_4$ | 7.64 | -0.94 | 0.68 | 0.70 |
| $P_5$ | $2.48 \cdot 10^{-6}$ | 0 | 1.44 | 0.66 |
| $P_6$ | $1.09 \cdot 10^{-5}$ | 0.04 | 1.00 | 1.04 |
| $P_7$ | $1.29 \cdot 10^{-1}$ | 0 | 0 | 0.94 |

A second model was created by fitting the GMA model (Eq. 5) against all five datasets (case 2.2, GMA (I-V), see file 'SWG_Case2_GMA.I_V.m'). The model did fit well datasets III-V, yet this came at the expense of a loss of fitness towards dataset II (Fig S7).

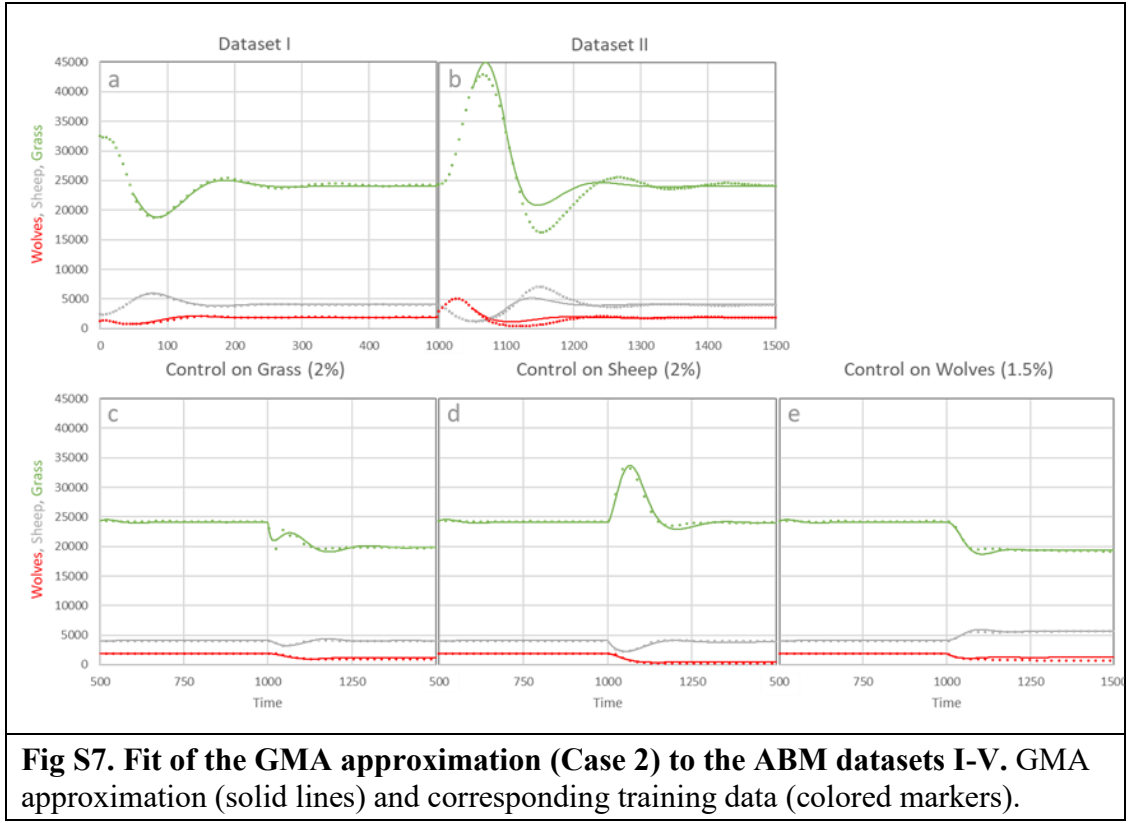

**Fig S7. Fit of the GMA approximation (Case 2) to the ABM datasets I-V.** GMA approximation (solid lines) and corresponding training data (colored markers).

**Step 6:** Both models were used to estimate the levels of control (in $\kappa_2$ and $\kappa_3$) that need to be exerted in order to reduce the steady state of wolves to 50% and increase sheep to 110% (Fig S7). Clearly the model parameterized against all 5 datasets (GMA (I-V)) performed better, predicting a control solution much closer to the ABM solution than when parameterized only against datasets I and II (GMA (I-V)).

## Case 3 Approximations in the vicinity of the steady state

Generation of dynamic approximations are more complex and require the inference of the mechanisms involved in the ABM. However, when an ABM has a stable steady state of interest, and the control problem will not take the model too far away from this steady state, then it is possible to generate surrogate models anchored at the steady state. We study steady-state stability in ODEs by linear approximation around the steady state. In this context, we propose applying a similar approach to a given ABM. The difference is that the elements of the Jacobian matrix are inferred by optimization rather than analytically calculated from the derivatives of the ODEs.

**Step 1:** Identification of the state variables. The state variables have been defined in earlier cases. For consistency in the comparison, the same grass, sheep, and wolves will be used. We created two approximations: a linear approximation

$$\frac{d}{dt}\begin{bmatrix} X \\ Y \\ Z \end{bmatrix} = J \cdot \left( \begin{bmatrix} X \\ Y \\ Z \end{bmatrix} - \begin{bmatrix} X_{SS} \\ Y_{SS} \\ Z_{SS} \end{bmatrix} \right), \qquad \text{(Eq. 7)}$$

and a quadratic approximation

$$\frac{d}{dt}\begin{bmatrix} X \\ Y \\ Z \end{bmatrix} = J \cdot \begin{bmatrix} X - X_{SS} \\ Y - Y_{SS} \\ Z - Z_{ss} \end{bmatrix} + H \cdot \begin{bmatrix} (X - X_{SS})^2 \\ (Y - Y_{SS})^2 \\ (Z - Z_{ss})^2 \\ (X - X_{SS}) \cdot (Y - Y_{SS}) \\ (X - X_{SS}) \cdot (Z - Z_{ss}) \\ (Y - Y_{SS}) \cdot (Z - Z_{ss}) \end{bmatrix}, \qquad \text{(Eq. 8)}$$

where $J$ is a 3×3 matrix of the first-order elements, and $H$ is a 3×6 matrix of the second-order elements, but not a true Hessian matrix.

**Step 2:** Determination of the steady state point. Using the datasets I and II and previous analyses, the steady state is $(X_{SS}, Y_{SS}, Z_{SS}) = (24,290, 4,001, 1,917)$.

**Step 3:** The control terms were approximated as in case 1, and the two models (Eq. 7 and 8) with the control terms are

$$\frac{d}{dt}\begin{bmatrix} X \\ Y \\ Z \end{bmatrix} = J \cdot \left( \begin{bmatrix} X \\ Y \\ Z \end{bmatrix} - \begin{bmatrix} X_{SS} \\ Y_{SS} \\ Z_{ss} \end{bmatrix} \right) - \begin{bmatrix} 0 \\ \kappa_2 \cdot Y \\ \kappa_3 \cdot Z \end{bmatrix}, \text{ and} \qquad \text{(Eq. 9)}$$

$$\frac{d}{dt}\begin{bmatrix} X \\ Y \\ Z \end{bmatrix} = J \cdot \begin{bmatrix} X - X_{SS} \\ Y - Y_{SS} \\ Z - Z_{ss} \end{bmatrix} + H \cdot \begin{bmatrix} (X - X_{SS})^2 \\ (Y - Y_{SS})^2 \\ (Z - Z_{ss})^2 \\ (X - X_{SS}) \cdot (Y - Y_{SS}) \\ (X - X_{SS}) \cdot (Z - Z_{ss}) \\ (Y - Y_{SS}) \cdot (Z - Z_{ss}) \end{bmatrix} - \begin{bmatrix} 0 \\ \kappa_2 \cdot Y \\ \kappa_3 \cdot Z \end{bmatrix}. \qquad \text{(Eq. 10)}$$

**Step 4:** Parameterization of the models. When we attempted to fit the linear model (Eq. 7) to the datasets I and II, a good solution could not be found. We then aimed to fit it against each of the two datasets and found better fits, although the fit against dataset I was better (Fig S8, see file 'SWG_Case3_Linear.I.m'). This underscores a limitation of this approach—it may not be able to approximate values that are too far away from the steady state.

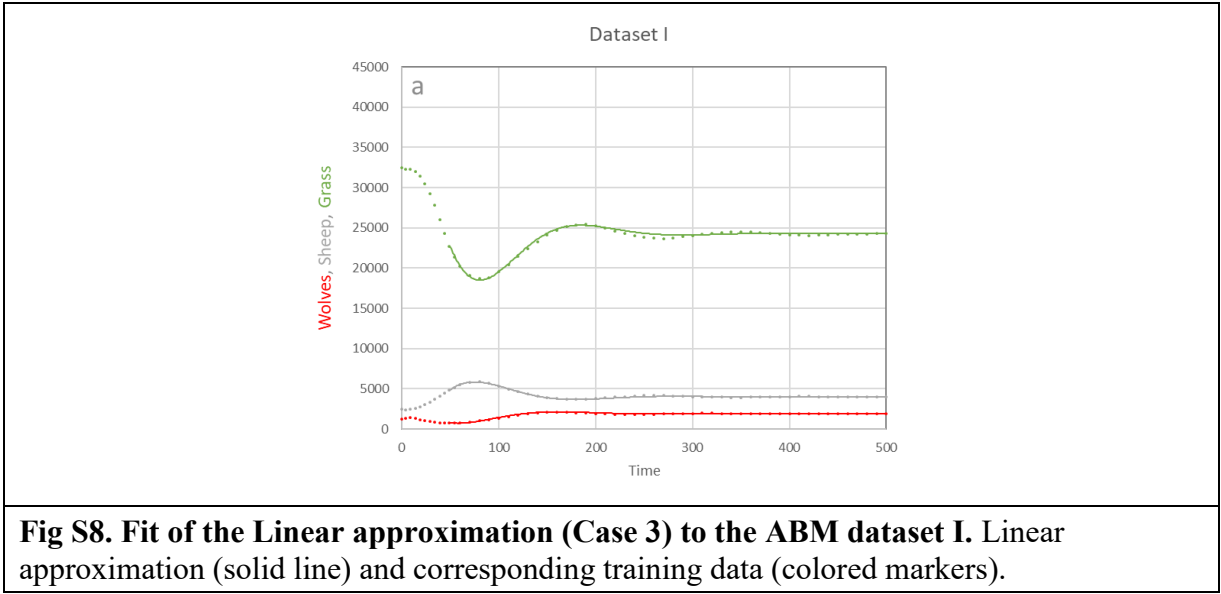

**Fig S8. Fit of the Linear approximation (Case 3) to the ABM dataset I.** Linear approximation (solid line) and corresponding training data (colored markers).

The quadratic model (Eq. 8), however, was able to fit both datasets I and II (see file 'SWG_Case3_Quad.I_II.m') and datasets I-V(see file 'SWG_Case3_Quad.I_V.m'). This is not surprising as this model (Eq. 8) has 27 parameters. For fitting datasets I-V, the system below was used (Eq. 11). Both optimizations generated good fits. However, when these models were used to predict the solution of the control problem, they did not perform well. The optimizations were unstable and did not converge. We recognized that second-order systems, unlike first-order systems, have more steady states and some of these are unstable. Thus, we re-parameterized the models against datasets I and II (Fig S9) and I-V (Fig S10), while excluding any parameterization that had extra steady states within the region of interest, and with this strategy it was possible to easily predict the solutions to the control problem.

$$\frac{d}{dt}\begin{bmatrix} X \\ Y \\ Z \end{bmatrix} = J \cdot \begin{bmatrix} X - X_{SS} \\ Y - Y_{SS} \\ Z - Z_{ss} \end{bmatrix} + H \cdot \begin{bmatrix} (X - X_{SS})^2 \\ (Y - Y_{SS})^2 \\ (Z - Z_{ss})^2 \\ (X - X_{SS}) \cdot (Y - Y_{SS}) \\ (X - X_{SS}) \cdot (Z - Z_{ss}) \\ (Y - Y_{SS}) \cdot (Z - Z_{ss}) \end{bmatrix} - \begin{bmatrix} \kappa_1 \cdot X \\ \kappa_2 \cdot Y \\ \kappa_3 \cdot Z \end{bmatrix} \qquad \text{(Eq. 11)}$$

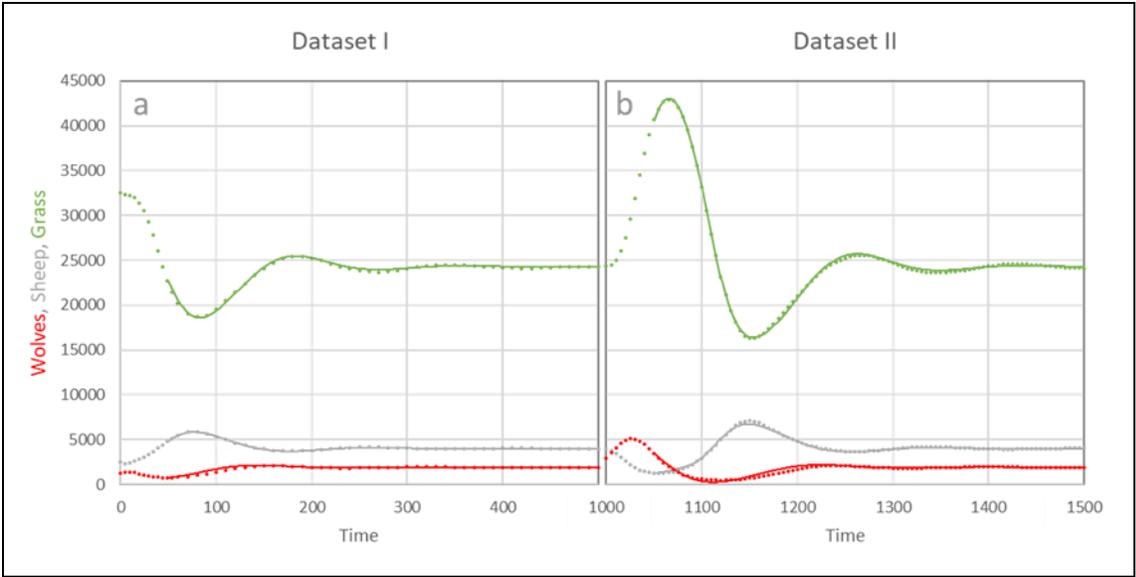

**Fig S9. Fit of the Quadratic approximation (Case 3) to the ABM datasets I and II.** Quadratic approximation (solid lines) and corresponding training data (colored markers).

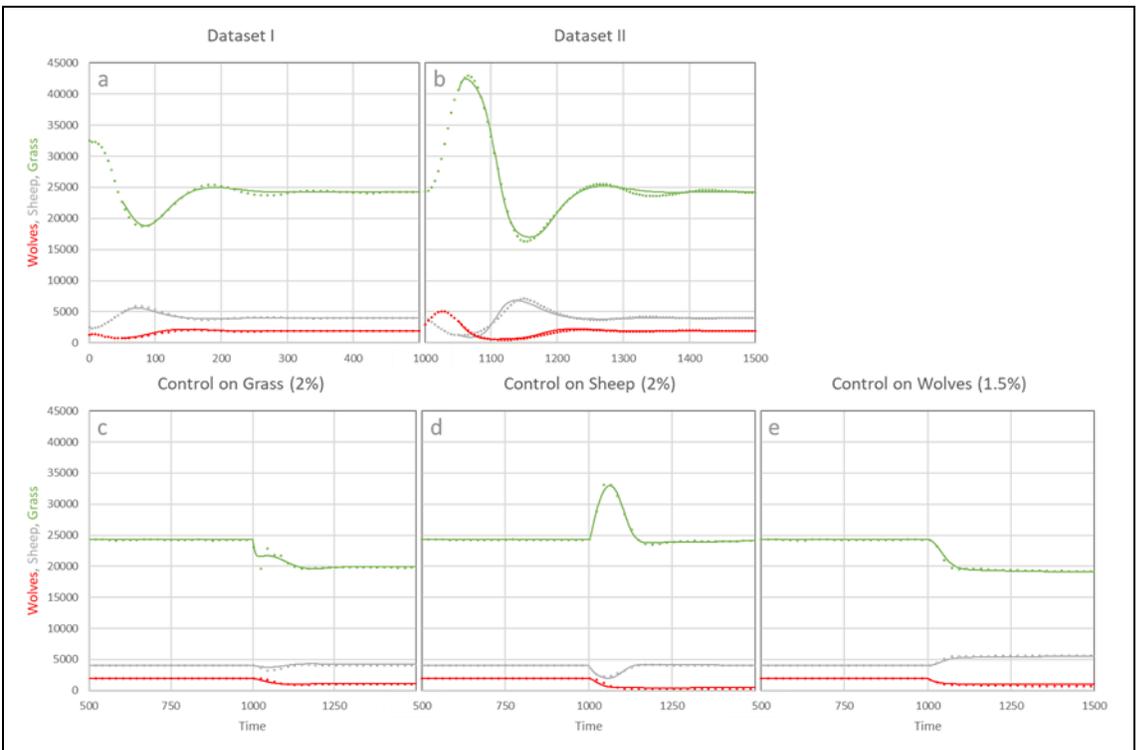

**Fig S10. Fit of the Quadratic approximation (Case 3) to the ABM datasets I-V.** Quadratic approximation (solid lines) and corresponding training data (colored markers).

**Step 5:** All three models (linear I and II, quadratic I and II, and quadratic I-V) were used to estimate the levels of control (in $\kappa_2$ and $\kappa_3$) needed to reduce the steady state of wolves to 50% and increase sheep to 110% (Fig S5). The linear approximation predicted a control solution $(\kappa_2, \kappa_3) = (0.0884, 0.0297)$ that was far from all other approaches and it is not shown in Fig S5. Both parameterizations of the quadratic approach are shown in Fig S5, and similarly to case 2 the parameterization obtained from all five datasets (I-V) predicted a much better control solution than the parameterization obtained from datasets I and II.

**Case 4 The S-system approach**
An S-system model allows one to approximate an ABM as a set of ODEs where each state variable is modeled as the difference between two power law terms (Eq. 12). The first term approximates all incoming processes, while the second approximates all outgoing processes.

**Step 1:** Identification of all state variables. We will use the same three state variables as in the previous cases: grass, sheep, and wolves. Therefore, the ODE system for a 3-variable S-system is

$$\dot{X} = \alpha_1 \cdot X^{g_{11}} \cdot Y^{g_{12}} \cdot Z^{g_{13}} - \beta_1 \cdot X^{h_{11}} \cdot Y^{h_{12}} \cdot Z^{h_{13}}$$
$$\dot{Y} = \alpha_2 \cdot X^{g_{21}} \cdot Y^{g_{22}} \cdot Z^{g_{23}} - \beta_2 \cdot X^{h_{21}} \cdot Y^{h_{22}} \cdot Z^{h_{23}}. \qquad \text{(Eq. 12)}$$
$$\dot{Z} = \alpha_3 \cdot X^{g_{31}} \cdot Y^{g_{32}} \cdot Z^{g_{33}} - \beta_3 \cdot X^{h_{31}} \cdot Y^{h_{32}} \cdot Z^{h_{33}}$$

**Step 2:** The control terms were approximated as in previous cases. The S-system model with the control terms is

$$\dot{X} = \alpha_1 \cdot X^{g_{11}} \cdot Y^{g_{12}} \cdot Z^{g_{13}} - \beta_1 \cdot X^{h_{11}} \cdot Y^{h_{12}} \cdot Z^{h_{13}}$$
$$\dot{Y} = \alpha_2 \cdot X^{g_{21}} \cdot Y^{g_{22}} \cdot Z^{g_{23}} - \beta_2 \cdot X^{h_{21}} \cdot Y^{h_{22}} \cdot Z^{h_{23}} - \kappa_2 \cdot Y. \qquad \text{(Eq. 13)}$$
$$\dot{Z} = \alpha_3 \cdot X^{g_{31}} \cdot Y^{g_{32}} \cdot Z^{g_{33}} - \beta_3 \cdot X^{h_{31}} \cdot Y^{h_{32}} \cdot Z^{h_{33}} - \kappa_3 \cdot Z$$

**Step 3:** Parameterization of the model. Similar to what was done previously, two parameterizations were generated: one against datasets I and II (Fig S11, see file 'SWG_Case4_Ssystem.I_II.m'), and a second against all five datasets I-V (Fig S12, see file 'SWG_Case4_Ssystem.I_V.m'). For the parameterization against all five datasets, we used the ODE system

$$\dot{X} = \alpha_1 \cdot X^{g_{11}} \cdot Y^{g_{12}} \cdot Z^{g_{13}} - \beta_1 \cdot X^{h_{11}} \cdot Y^{h_{12}} \cdot Z^{h_{13}} - \kappa_1 \cdot X$$
$$\dot{Y} = \alpha_2 \cdot X^{g_{21}} \cdot Y^{g_{22}} \cdot Z^{g_{23}} - \beta_2 \cdot X^{h_{21}} \cdot Y^{h_{22}} \cdot Z^{h_{23}} - \kappa_2 \cdot Y. \qquad \text{(Eq. 14)}$$
$$\dot{Z} = \alpha_3 \cdot X^{g_{31}} \cdot Y^{g_{32}} \cdot Z^{g_{33}} - \beta_3 \cdot X^{h_{31}} \cdot Y^{h_{32}} \cdot Z^{h_{33}} - \kappa_3 \cdot Z$$

Both optimizations produced good fits (Figs S11 and S12). Similar to cases 2 and 3, the fit was slightly less accurate when considering all five datasets compared to when considering just datasets I and II.

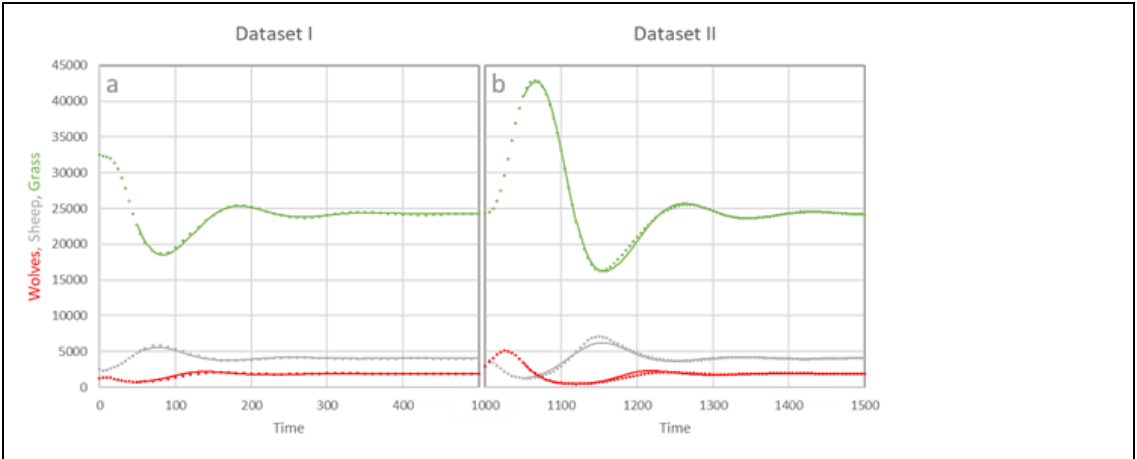

**Fig S11. Fit of the S-system approximation (Case 4) to the ABM datasets I and II.** The S-system approximation (solid line) and corresponding training data (colored markers).

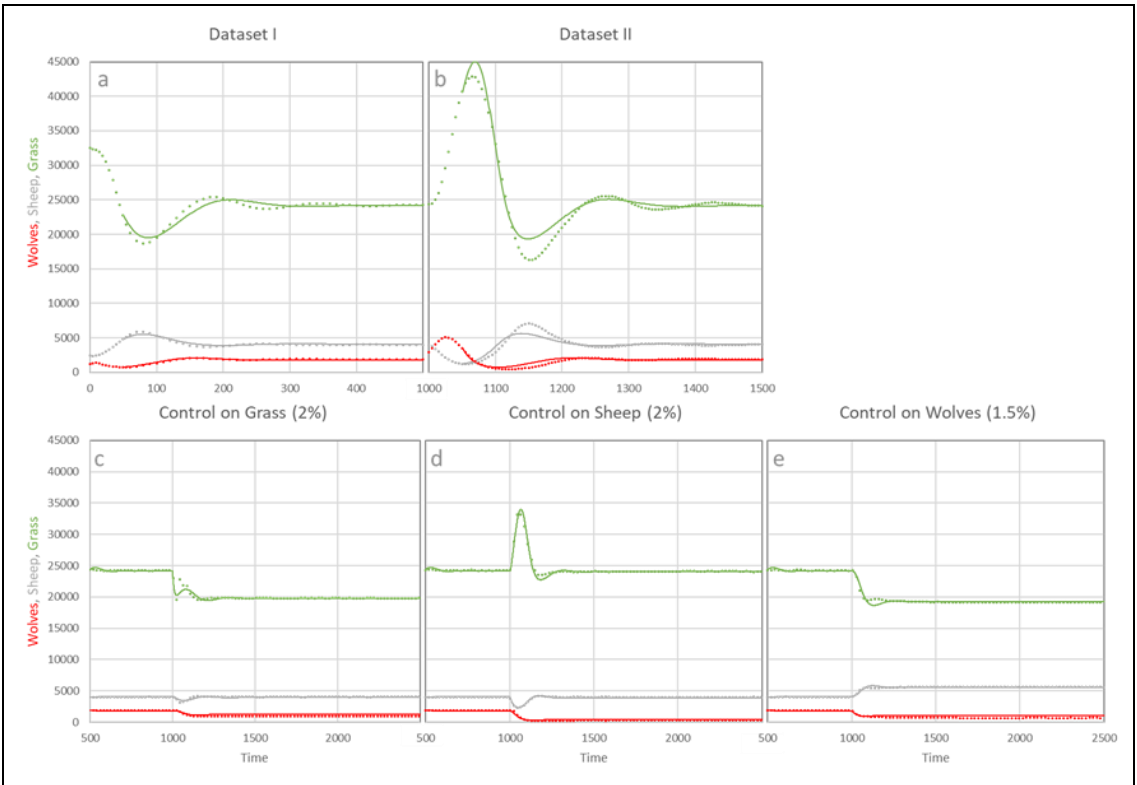

**Fig S12. Fit of the S-system approximation (Case 4) to the ABM datasets I-V.** The S-system approximation (solid line) and corresponding training data (colored markers).

**Step 4:** Both parameterizations were used to estimate the levels of control (in $\kappa_2$ and $\kappa_3$) needed to reduce the steady state of wolves to 50% and increase sheep to 110% (Fig S5). Both parameterizations of the S-system approach are shown in Fig S5, and similarly to cases 2 and 3 the parameterization obtained from all five datasets predicted a much better control solution than the parameterization obtained from datasets I and II.

## The metabolic pathway model

In the first example, we considered the sheep-wolves-grass model, and trained all approximations with control data to create surrogate models. We could not identify significant differences in the ability of the considered surrogates to determine the optimal control solution. To provide further insights into the different approximations, we developed a simplified metabolic ABM. This ABM has a mechanistic approximation based on Michaelis–Menten kinetics. The use of this kinetics adds complexity, as accurately approximating Michaelis–Menten kinetics over a wide range of substrate concentrations is challenging for power laws. The pathway deliberately includes divergent processes as these are difficult to be captured accurately with S-system approximations.

The metabolic model (Fig S13) is based on four reactions catalyzed by four enzymes (A, E, I, and O), and five metabolites (S, P, Q, R, and T). Enzyme A, catalyzes the conversion of S into P, and is competitively inhibited by R, enzyme E, catalyzes the conversion of P into Q, enzyme I, catalyzes the conversion of Q into R, and enzyme O, catalyzes the conversion of P into T and is activated by R (Figs S13A and S13B). Activation of the conversion of P into T by R is modeled by setting the rate of conversion of [ORP] into [ORT] faster than the rate of conversion of [OP] into [OT] (Fig S13B). Enzyme A forms complexes AS, AP and AR, enzyme E, EP and EQ, enzyme I, IQ and IR, and enzyme O, OP, OT, OR, ORP, and ORT (Fig S13B). The model assumes all metabolites move 10 times faster than enzymes and complexes, and all species move randomly in a continuous space of 100×100 with periodic boundary conditions. A metabolite is available to bind with an enzyme or complex when they are at the same grid point, modeled by flooring (*floor* function) their positions. Two types of simulations were used, batch (Fig S13A) and continuous (Fig S13C). In batch mode, the simulations were started with 200 agents of each enzyme, 80,000 of S, 20,000 of P, 20,000 of Q, 10 of R, 10 of T, and no enzymatic complexes were present initially (Fig S14). In continuous mode, the simulations were started with the same initial conditions, but there was a constant inflow of S, at a rate of one agent per timestep and a constant removal of all metabolites, at a rate of 0.05% per timestep (Fig S15).

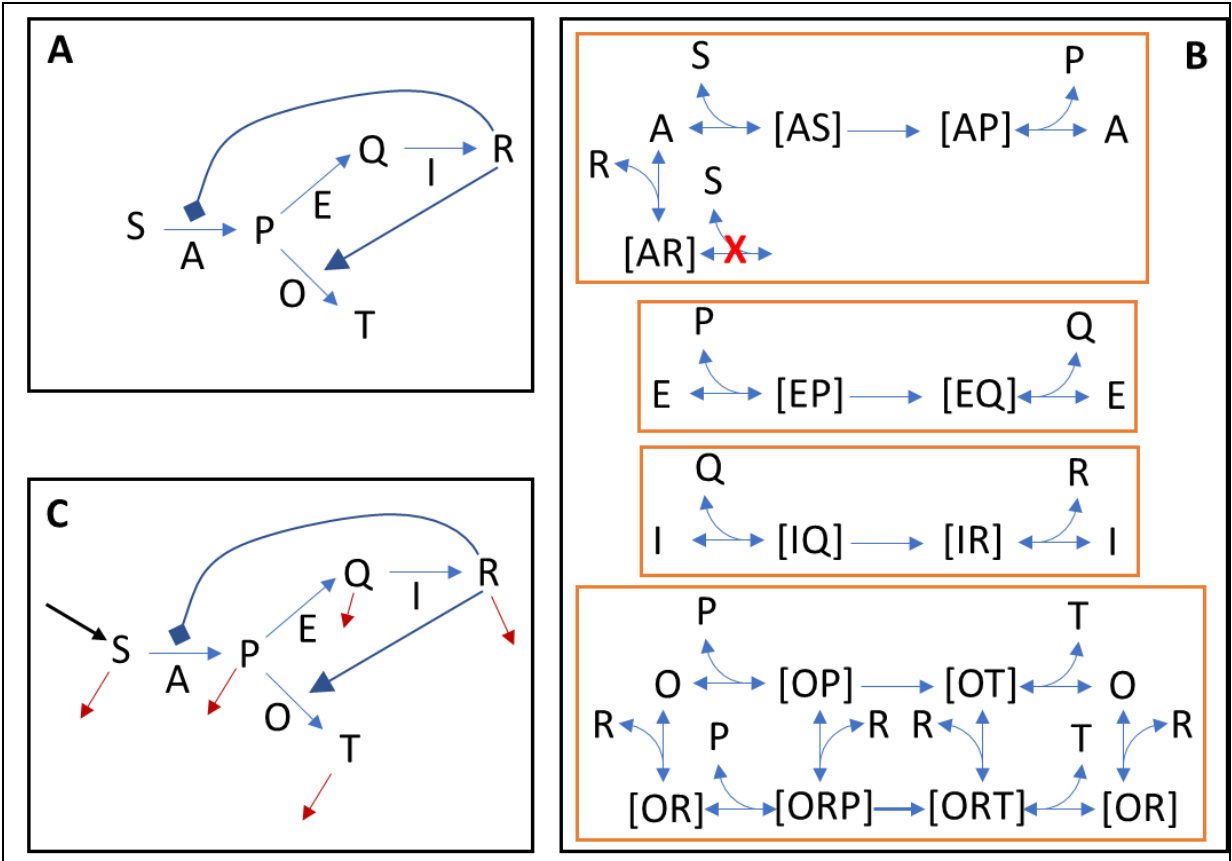

**Fig S13. Metabolic toy ABM representations.** (A) The diagram of the macroscopic representation of the ABM corresponding to the microscopic diagram in (B). (C) The macroscopic representation when the model is used in continuous mode, where a constant inflow of agents S occurs while all metabolites are removed at a constant rate. In (B), all pairwise interactions and complex decompositions are modeled with different probabilities. Two agents are able to interact when present in the same grid point, which was modeled with a *floor* function.

The control problem that we wish to solve is to determine the constant rate of inflow of S that minimizes the waist of S and maximizes the production of R and T, while all metabolites are removed at a rate of 0.05% per timestep. Mathematically, the goal is to identify the constant inflow, $Q_{in}$, that minimizes the loss function,

$$Loss(Q_{in}) = \sum_{k=1}^{N_t} \frac{S_k}{R_k + T_k},$$ (Eq. 15)

during a simulation run of $N_t$=50,000 timesteps, where $S_k$, $R_k$ and $T_k$ are the concentrations of S, R and T at timestep $k$.

To compare the ability of all proposed ODE approximations (mechanistic, GMA, S-system, quadratic, and linear), we optimized them against the two collections of datasets ('I' will denote models optimized against datasets I and II, generated with a single simulation of each condition; and 'C' will denote models optimized against datasets III-V, generated by averaging 100 simulations of each condition). Dataset I (generated with file 'Met_Pathway_dataset_80k_20k_20k_10_10_NoDil.m') was obtained as a single simulation of the metabolic ABM with the initial conditions listed above and under batch mode (Fig S14), and dataset II (generated with file 'Met_Pathway_dataset_80k_20k_20k_10_10_wDil_wFeed.m') was obtained as a single simulation of the metabolic ABM with the initial conditions listed above, under continuous mode, with an inflow rate of 1 agent of S per timestep and a removal rate of 0.05% per timestep (Fig S14). Dataset III (Fig S14) was obtained by averaging 100 simulations started in the same initial conditions as dataset I, and dataset IV (Fig S14) was obtained by averaging 100 simulations started using the same initial conditions and continuous mode as dataset II. Finally, dataset V was obtained by averaging 100 simulations started using the same initial conditions as dataset IV, and in continuous mode but with an inflow of 0.2 agents of S per timestep (Fig S14).

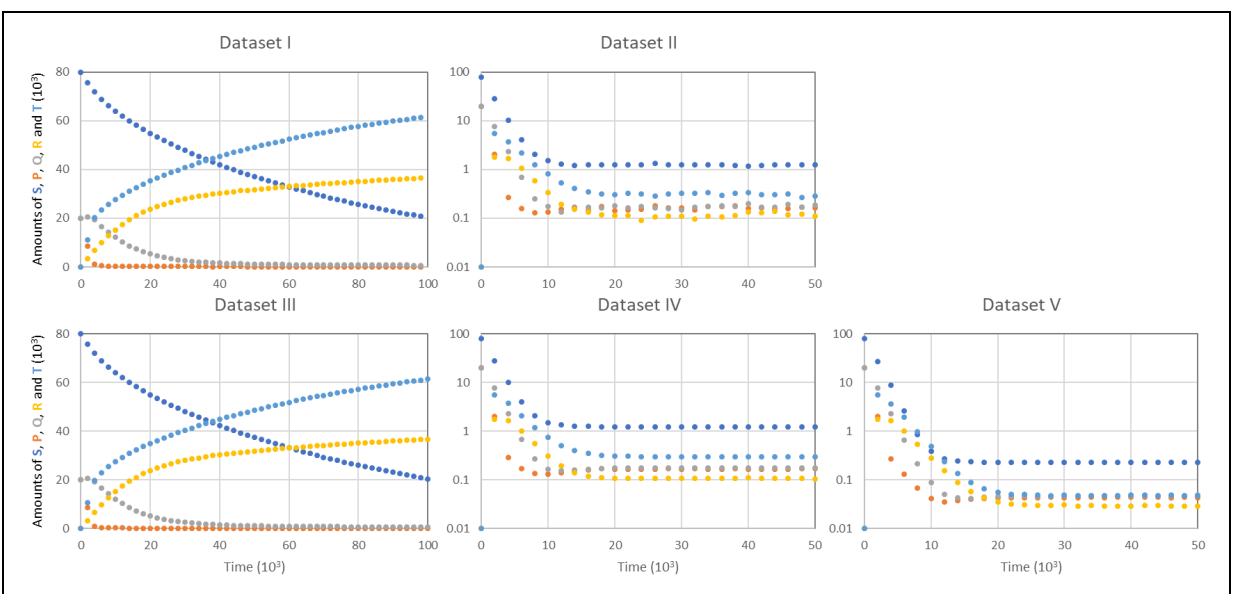

**Fig S14. Datasets used to train the surrogate models for the metabolic ABM.** Collection I is made of datasets I&II, and collection C is made of datasets III-V. For visualization purposes the datasets were down sampled from 1 point per timestep to 1 point per 2,000 timesteps. All calculations were performed using 1 point per timestep.

## Case 1 - Mechanistic

For the mechanistic approximation each of the microscopic reactions depicted in Fig S13B were approximated to a Michaelis–Menten type rate law, which reduced the ODE model to the diagram shown in Fig S13A. The reaction catalyzed by A was approximated as a single-substrate irreversible reaction with two competitive inhibitors, P and R (Eq. 16). The reactions catalyzed by E and I were approximated as single-substrate irreversible reactions with product inhibition (Eq. 16), and the reaction catalyzed by O was approximated as a mixed noncompetitive activation (also known as heterotropic allosteric activation), where the conversion of P to T occurs faster when R is bound to O (Eq. 16). The ODE approximation is given by

$$\frac{dX}{dt} = \begin{bmatrix} -1 & 0 & 0 & 0 \\ 1 & -1 & 0 & -1 \\ 0 & 1 & -1 & 0 \\ 0 & 0 & 1 & 0 \\ 0 & 0 & 0 & 1 \end{bmatrix} \cdot \begin{bmatrix} F_A \\ F_E \\ F_I \\ F_O \end{bmatrix} + \begin{bmatrix} Q_{in} \\ 0 \\ 0 \\ 0 \\ 0 \end{bmatrix} - k_{out} \cdot X,$$

$$F_A = \frac{p_1 \cdot \frac{X_1}{p_2}}{1 + \frac{X_1}{p_2} + \frac{X_2}{p_3} + \frac{X_4}{p_4}}, F_E = \frac{p_5 \cdot \frac{X_2}{p_6}}{1 + \frac{X_2}{p_6} + \frac{X_3}{p_7}}, F_I = \frac{p_8 \cdot \frac{X_3}{p_9}}{1 + \frac{X_3}{p_9} + \frac{X_4}{p_{10}}},$$  (Eq. 16)

$$F_O = \frac{p_{11} \cdot \frac{X_2}{p_{12}} + p_{13} \cdot \frac{X_4}{p_{14}} \cdot \frac{X_2}{p_{15}}}{1 + \frac{X_2}{p_{12}} + \frac{X_5}{p_{16}} + \frac{X_4}{p_{14}} \cdot \left(1 + \frac{X_2}{p_{15}} + \frac{X_5}{p_{17}}\right)},$$

where $X \in \mathbb{R}^5$ is the vector of the state variables representing the concentrations of S, P, Q, R and T, respectively; $F_A$, $F_E$, $F_I$, and $F_O$ are the fluxes through the reactions catalyzed by A, E, I, and O, respectively; $Q_{in}$ is the inflow of S when the system is in continuous mode; $k_{out}$ is the rate of removal of metabolites when the system is in continuous mode; and $p_j \in \mathbb{R}^{17}$ is the vector of kinetic parameters.

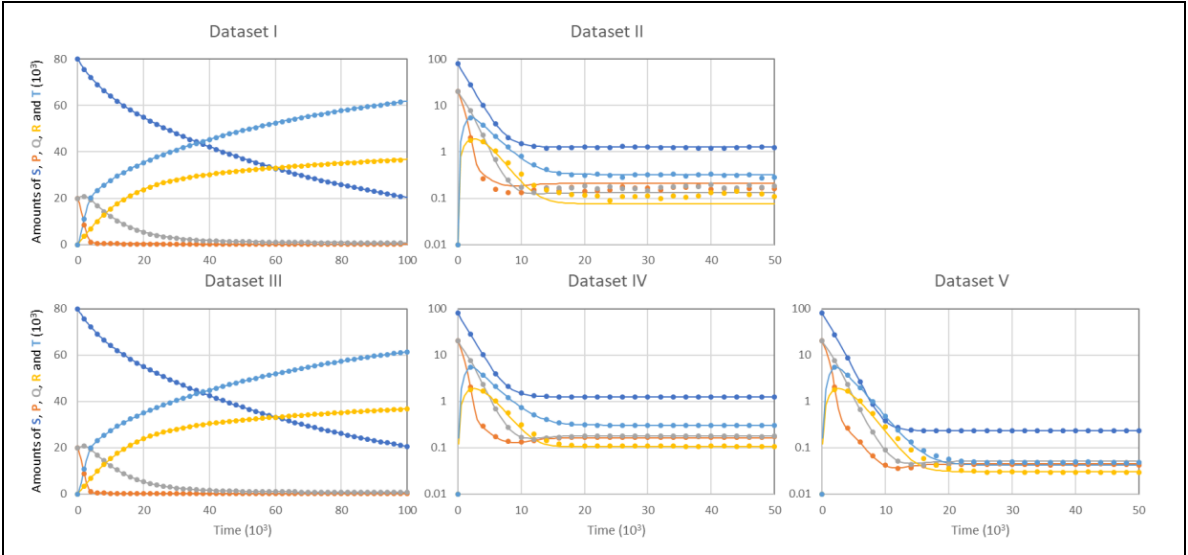

**Fig S15. Fit of the mechanistic approximation (Case 1) to the ABM dataset collections I and C.** The mechanistic approximation (solid line) and corresponding training data (colored markers). The Mech I model fit was obtained with datasets I&II (collection I) and the Mech C model fit was obtained with datasets III-V (collection C). For visualization purposes the datasets were down sampled from 1 point per timestep to 1 point per 2,000 timesteps.

This mechanistic ODE model was parameterized against the two collections of datasets (Fig S15), which resulted in two models: Mech. I(see file 'MetPw_Case1_Mech.I.m') and Mech. C(see file 'MetPw_Case1_Mech.C.m'). These two models were then evaluated for their ability to identify the solution to the control problem (Eq. 15), and the results are shown in Fig S16. The solution of the control problem was also directly evaluated on the ABM by grid search in the domain of $Q_{in}$ between 0 and 1 with a step size on 0.1. At each evaluation point, 100 simulations were averaged (Fig S16). The optimal value found for the ABM was $Q_{in}$=0.7. Both parameterizations of the mechanistic model were good at predicting the optimal loss function value and optimal inflow of S.

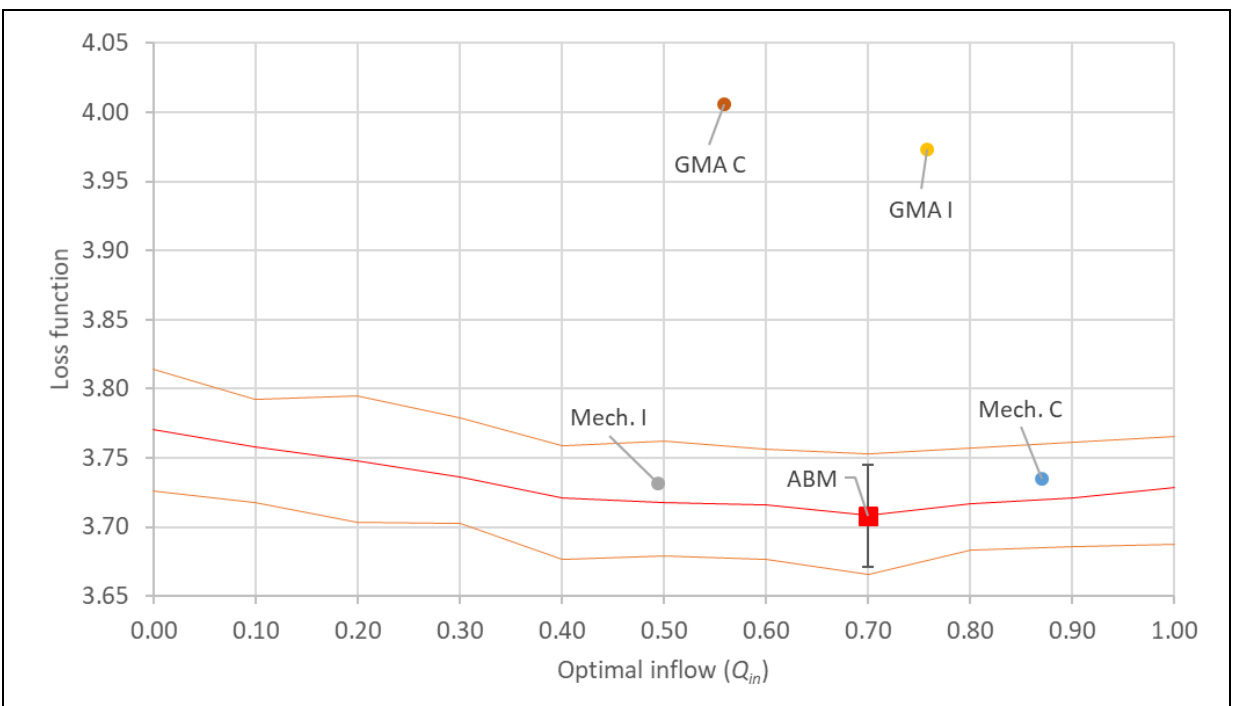

**Fig S16. Comparison of effectiveness of different ODE surrogates for solving the metabolic pathway ABM control problem.** The red square shows the optimal inflow point and the corresponding mean loss function value as determined for the ABM by a grid search between 0 and 1.0 with a step size of 0.1, where in each step 100 simulations runs were averaged. The red line highlights the mean of each of the 100 simulation runs of the ABM and the orange line the 75% confidence band. Circles denote the predicted optimal inflow and corresponding loss function value for each ODE surrogate. GMA I was the surrogate that best predicted an optimal inflow of substrate closest to the ABM and Mech. I was best at predicting the loss function value of the ABM at the optimal inflow point.

**Case 2 – GMA approximation**

For the GMA approximation of the metabolic ABM, we used the same stoichiometric matrix as in the mechanistic approximation. On the other hand, the fluxes through each enzyme were approximated to power law terms of all metabolites, and no assumptions were made about which metabolites regulate each of the fluxes. The GMA ODE model is

$$\frac{dX}{dt} = \begin{bmatrix} -1 & 0 & 0 & 0 \\ 1 & -1 & 0 & -1 \\ 0 & 1 & -1 & 0 \\ 0 & 0 & 1 & 0 \\ 0 & 0 & 0 & 1 \end{bmatrix} \cdot \begin{bmatrix} F_A \\ F_E \\ F_I \\ F_O \end{bmatrix} + \begin{bmatrix} Q_{in} \\ 0 \\ 0 \\ 0 \\ 0 \end{bmatrix} - k_{out} \cdot X,$$

$$F_A = p_1 \cdot X_1^{p_5} \cdot X_2^{p_9} \cdot X_3^{p_{13}} \cdot X_4^{p_{17}} \cdot X_5^{p_{21}},$$
$$F_E = p_2 \cdot X_1^{p_6} \cdot X_2^{p_{10}} \cdot X_3^{p_{14}} \cdot X_4^{p_{18}} \cdot X_5^{p_{22}},$$
$$F_I = p_3 \cdot X_1^{p_7} \cdot X_2^{p_{11}} \cdot X_3^{p_{15}} \cdot X_4^{p_{19}} \cdot X_5^{p_{23}},$$
$$F_O = p_4 \cdot X_1^{p_8} \cdot X_2^{p_{12}} \cdot X_3^{p_{16}} \cdot X_4^{p_{20}} \cdot X_5^{p_{24}},$$

where $X \in \mathbb{R}^5$ is the vector of the state variables representing the concentrations of S, P, Q, R and T; $F_A$, $F_E$, $F_I$, and $F_O$ are the fluxes through the reactions catalyzed by A, E, I, and O, respectively; $Q_{in}$ is the inflow of S into the system when the system is in continuous mode; $k_{out}$ is the rate of removal of metabolites when the system is in continuous mode; and $p_j \in \mathbb{R}^{24}$ is the vector of kinetic parameters.

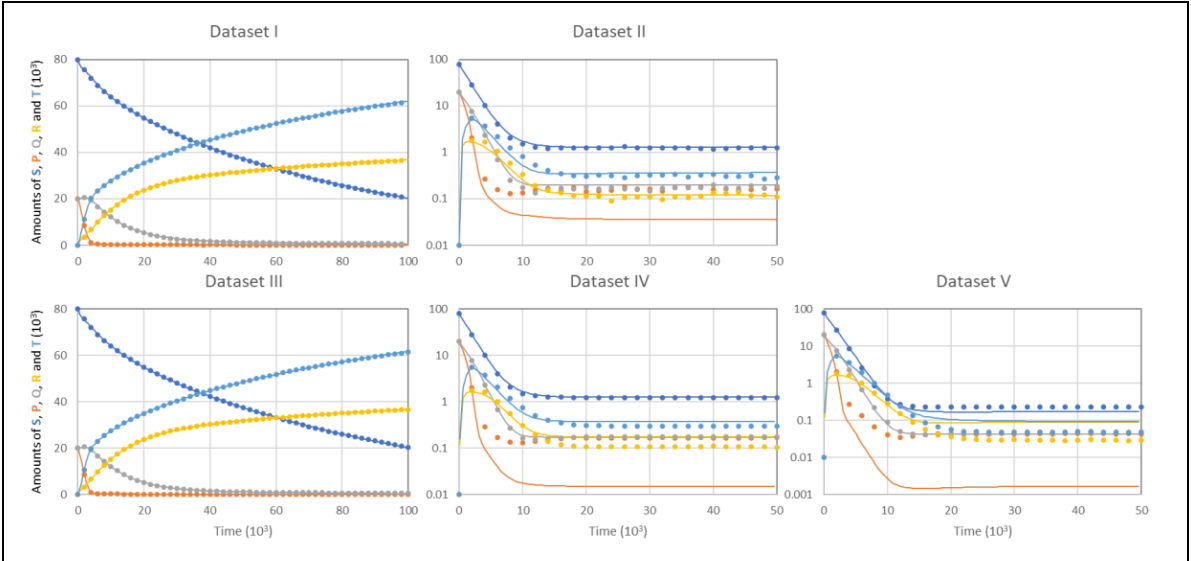

**Fig S17. Fit of the GMA approximation (Case 2) to the ABM dataset collections I and C.** The GMA approximation (solid line) and corresponding training data (colored markers). The GMA I model fit was obtained with datasets I and II (collection I) and the GMA C model fit was obtained with datasets III-V (collection C). For visualization purposes the datasets were down sampled from 1 point per timestep to 1 point per 2,000 timesteps.

The GMA model was parameterized against the two collections of datasets, I and C (Fig S17), which resulted in two models: GMA I(see file 'MetPw_Case2_GMA.I.m') and GMA C(see file 'MetPw_Case2_GMA.C.m'). These two models were then evaluated for their ability to predict the solution to the control problem (Eq. 15), and the results are shown in Fig S16. Both parameterizations of the GMA model identified similar optima for the inflow of S. The GMA solutions were better than the ones obtained with the mechanistic model, as they are closer to the true value determined for the ABM. Yet, the GMA models were not as good at identifying the optimal loss function value, and so, overall, the mechanistic approximations performed better than the GMA. Also, the fits of the mechanistic model (Fig S15) were better then the ones obtained for the GMA model (Fig S17).

**Case 3 Linear and quadratic approximations in the vicinity of the steady state**

We attempted to use the linear and quadratic approximations at the steady state to approximate the metabolic model, even though the model does not have an immediate steady state of interest. We therefore choose the trivial steady state of the dataset I (Fig S14). The steady state of dataset I occurs when the simulation is extended longer than the time horizon shown in Fig S14 and all of S, P, and Q get depleted and all of the mass of the system accumulates in R and T (*[S, P, Q, R, T]$_{SS}$=[0, 0, 0, 43226, 76794]*). The following ODE models were used:

$$\frac{dX}{dt} = J \cdot \overline{X} + \begin{bmatrix} Q_{in} \\ 0 \\ 0 \\ 0 \\ 0 \end{bmatrix} - k_{out} \cdot X, \qquad \text{(Eq. 18)}$$

$$\frac{dX}{dt} = J \cdot \overline{X} + H \cdot \overline{X2} + \begin{bmatrix} Q_{in} \\ 0 \\ 0 \\ 0 \\ 0 \end{bmatrix} - k_{out} \cdot X, \qquad \text{(Eq. 19)}$$

where $X \in \mathbb{R}^5$ is the vector of the state variables representing the concentrations of S, P, Q, R and T; $\overline{X}$=X-X$_{SS}$ is the vector of the centered first-order terms; $\overline{X2} = \begin{bmatrix} \overline{X}_1 \cdot \overline{X}_1, \overline{X}_1 \cdot \overline{X}_2, \overline{X}_1 \cdot \overline{X}_3, \overline{X}_1 \cdot \overline{X}_4, \overline{X}_1 \cdot \overline{X}_5, \overline{X}_2 \cdot \overline{X}_2, \overline{X}_2 \cdot \overline{X}_3, \overline{X}_2 \cdot \overline{X}_4, \overline{X}_2 \cdot \overline{X}_5, \overline{X}_3 \cdot \overline{X}_3, \overline{X}_3 \cdot \overline{X}_4, \overline{X}_3 \cdot \overline{X}_5, \overline{X}_4 \cdot \overline{X}_4, \overline{X}_4 \cdot \overline{X}_5, \overline{X}_5 \cdot \overline{X}_5 \end{bmatrix}^T$ is the vector of centered second-order terms; $J$ is the (5×5) Jacobian matrix of first-order parameters; H is the (5×15) matrix of second-order parameters; $Q_{in}$ is the inflow of S into the system when the system is in continuous mode; and $k_{out}$ is the rate of removal of metabolites when the system is in continuous mode.

Both approximations, Linear (Eq. 18) and Quadratic (Eq. 19), were parameterized against the datasets I and II, which resulted in two models: Linear I (Fig S18, see file 'MetPw_Case3_Linear.I.m'), and Quad I (Fig S19, see file 'MetPw_Case3_Quad.I.m'). These models were then evaluated for their ability to identify the solution to the control problem (Eq. 15), and the results are shown in Fig S20. None of the models were able to predict the optimal inflow of S, as none had a minimum between 0 and 1. The models displayed stiffness, making numerical integration challenging across the entire domain. Linear I could only be simulated for $Q_{in}$ values between 0.45 and 1, and Quad I for values between 0.4 and 1. Given these results, we did not attempt to fit these approximations to datasets III-V.

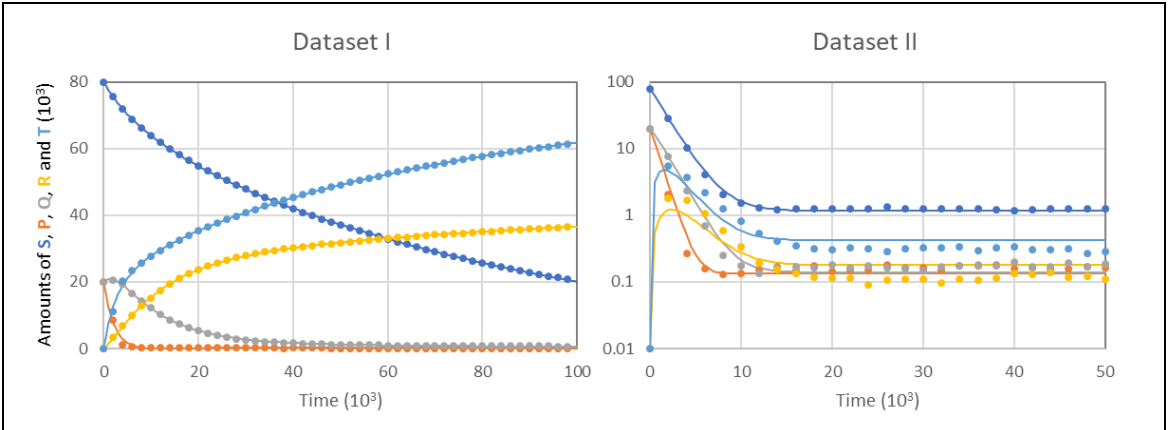

**Fig S18. Fit of the linear approximation (Case 3) to the ABM dataset collection I.**
The linear approximation (solid line) and corresponding training data (colored markers).
The Linear I model fit was obtained with datasets I and II (collection I). For
visualization purposes the datasets were down sampled from 1 point per timestep to 1
point per 2,000 timesteps.

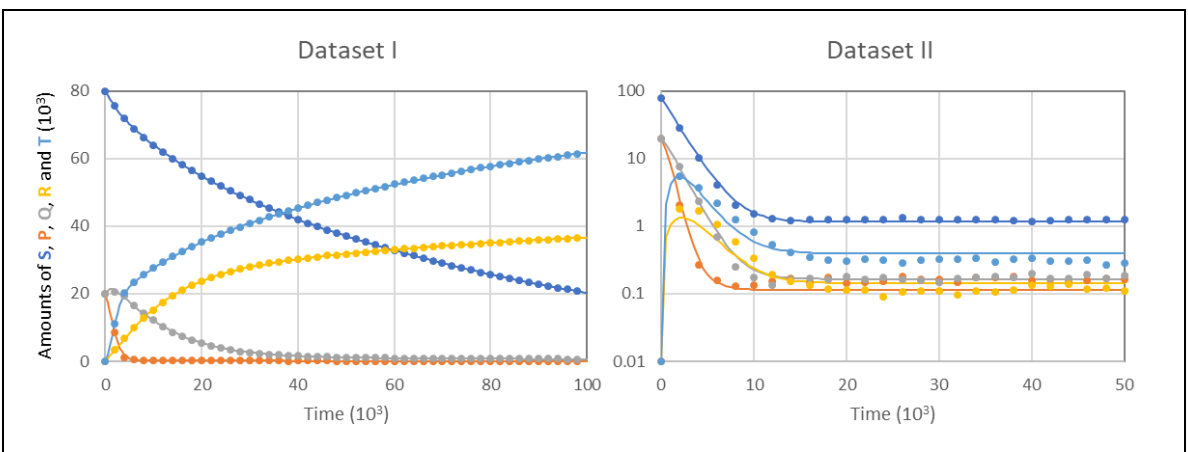

**Fig S19. Fit of the quadratic approximation (Case 3) to the ABM dataset collection I.**
The quadratic approximation (solid line) and corresponding training data (colored
markers). The Quad I model fit was obtained with datasets I and II (collection I). For
visualization purposes the datasets were down sampled from 1 point per timestep to 1
point per 2,000 timesteps.

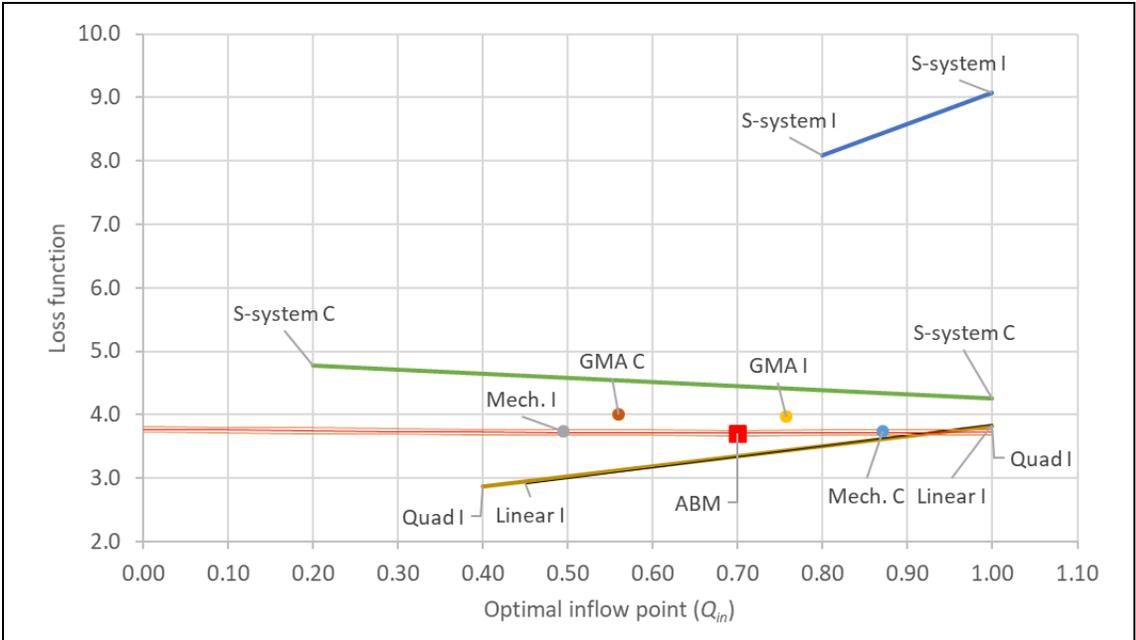

**Fig S20. Comparison of effectiveness of different ODE surrogates for solving the metabolic pathway ABM control problem.** The red square shows the optimal inflow point and the corresponding mean loss function value as determined for the ABM by a grid search between 0 and 1.0 with a step size of 0.1, where in each step 100 simulations runs were averaged. The red line highlights the mean of each of the 100 simulation runs of the ABM and the orange line the 75% confidence band. Circles denote the predicted optimal inflow and corresponding loss function value for each ODE surrogate. ODE models that did not exhibit a minimum within the 0 to 1.0 domain have their domain of integrability shown with a line. The line depicts the range of loss function values predicted by the approximation. The S-system I performed worst, as it could only be integrated between 0.8 and 1.0, and in that range predicted loss function values between 8 and 9. While S-system C, Quad I, and Linear I, all resulted in models with a larger domain over which they could be integrated, neither had a minimum within their respective domains. GMA I was the ODE surrogate that best predicted an optimal inflow of substrate closest to the ABM and Mech. I best predicted the loss function value of the ABM at the optimal inflow point.

**Case 4 The S-system approximation**

The S-system approximation of the metabolic ABM is given by the following ODE model:

$$\frac{dX}{dt} = \begin{bmatrix} \alpha_1 \cdot \prod_{j=1}^{5} X_j^{g_{1j}} - \beta_1 \cdot \prod_{j=1}^{5} X_j^{h_{1j}} \\ \alpha_2 \cdot \prod_{j=1}^{5} X_j^{g_{2j}} - \beta_2 \cdot \prod_{j=1}^{5} X_j^{h_{2j}} \\ \alpha_3 \cdot \prod_{j=1}^{5} X_j^{g_{3j}} - \beta_3 \cdot \prod_{j=1}^{5} X_j^{h_{3j}} \\ \alpha_4 \cdot \prod_{j=1}^{5} X_j^{g_{4j}} - \beta_4 \cdot \prod_{j=1}^{5} X_j^{h_{4j}} \\ \alpha_5 \cdot \prod_{j=1}^{5} X_j^{g_{5j}} - \beta_5 \cdot \prod_{j=1}^{5} X_j^{h_{5j}} \end{bmatrix} + \begin{bmatrix} Q_{in} \\ 0 \\ 0 \\ 0 \\ 0 \end{bmatrix} - k_{out} \cdot X, \qquad \text{(Eq. 20)}$$

where $X \in \mathbb{R}^5$ is the vector of the state variables representing the concentrations of S, P, Q, R and T, respectively; $Q_{in}$ the inflow of S into the system when the system is in continuous mode; $k_{out}$ is the rate of removal of metabolites when the system is in continuous mode; $\alpha_i$, $\beta_i \in \mathbb{R}^+$ are the rate constants; $g_{ij}$, $h_{ij} \in \mathbb{R}$ are the kinetic orders; and $i, j \in \{1,2,3,4,5\}$.

The S-system approximation was parameterized against the two collections of datasets, I and C (Fig S21), which resulted in two models: S-system I(see file 'MetPw_Case4_Ssystem.I.m') and S-system C (see file 'MetPw_Case4_Ssystem.C.m'). These two models were then evaluated for their ability to identify the solution to the control problem (Eq. 15), and the results are shown in Fig S20. Neither parameterizations of the S-system model were able to identify an optimal inflow of S as neither exhibit a loss function with a minimum between 0 and 1. The worst was S-system I as it was only able to be simulated between 0.8 and 1, while S-system C having been trained with datasets with $Q_{in}$ of 0.2 and 1 was able to be simulated within this range (*0.2<$Q_{in}$<1*). The failure of the S-system to approximate the metabolic ABM is not surprising as this ABM was specifically designed to be challenging for the S-system method. This ABM's processes are Michaelian, and S starts at high values relative to the Km of enzyme A. This poses an issue to power laws as these functions are only good at approximating Michaelis-Menten processes in small regions (Fig S22), which is likely what to be found if we were dealing with a real *in vivo* metabolic pathway. However, this same argument applies to the GMA approximation and this approximation did perform well. Additionally, the S-system also has problems approximating nodes with several processes coming in or out, as all processes in and all processes out of a node are approximated to single power laws. This is in contrast to the GMA approximation that has a power law function for each process and therefore does not suffer from this issue. Most likely, these two problems together prevented the S-system from producing good approximations to this ABM.

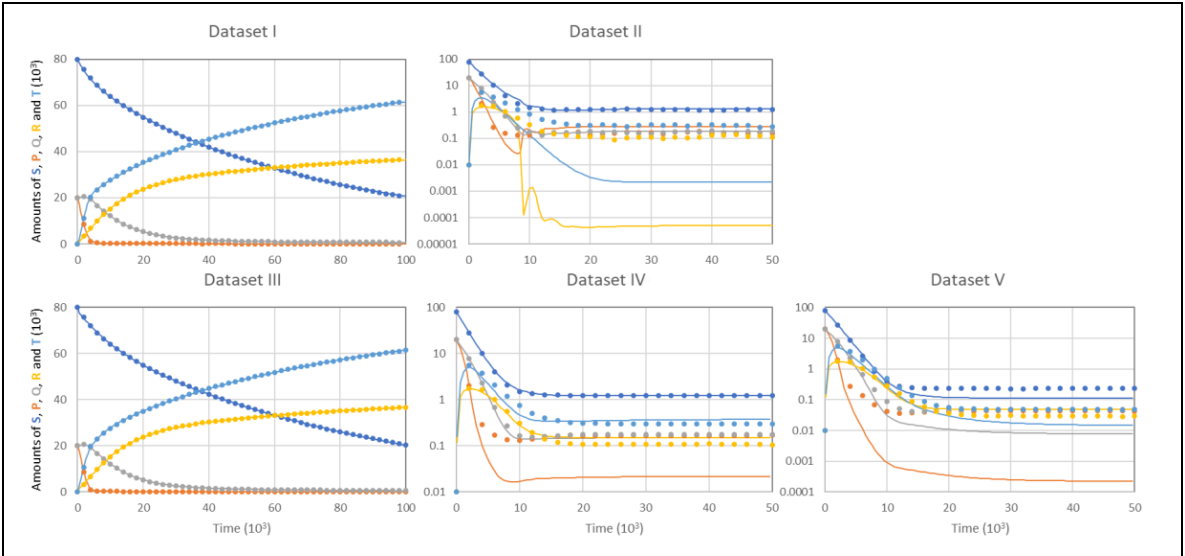

**Fig S21. Fit of the S-system approximation (Case 4) to the ABM dataset collections I and C.** The S-system approximation (solid line) and corresponding training data (colored markers). The S-system I model fit was obtained with datasets I and II (collection I) and the S-system C model fit was obtained with datasets III-V (collection C). For visualization purposes the datasets were down sampled from 1 point per timestep to 1 point per 2,000 timesteps.

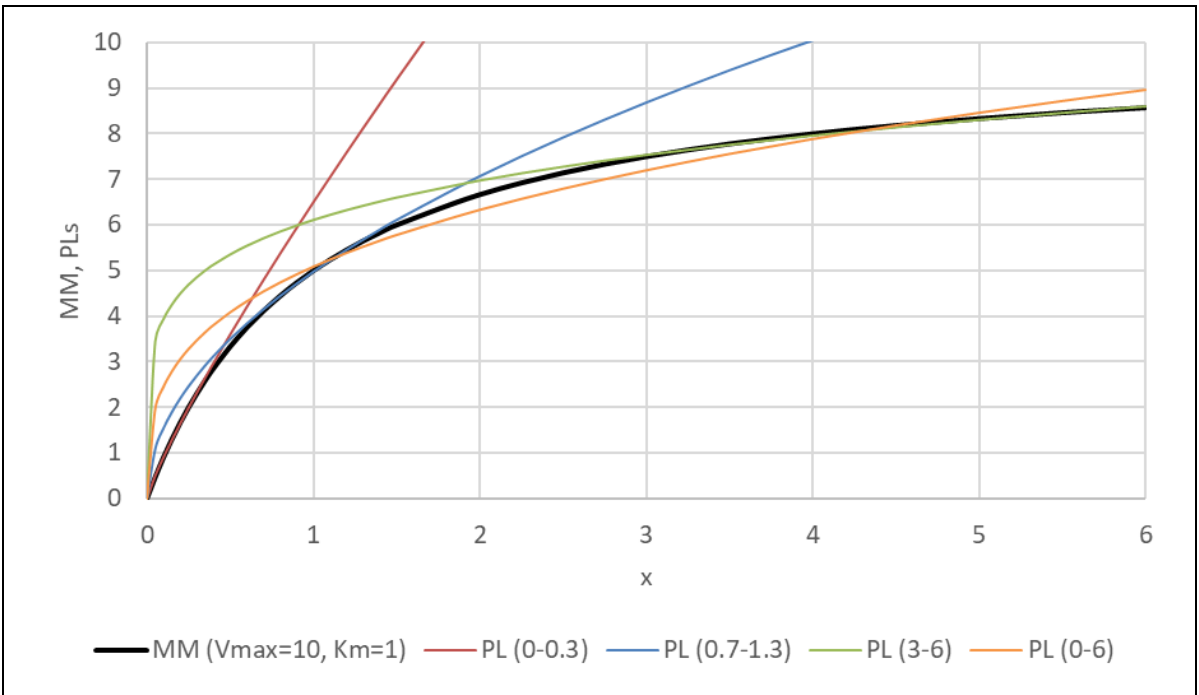

**Fig S22. Comparison of a Michaelis-Menten curve and 4 power laws fitted against different regions of the Michaelis-Menten curve.** The solid black line shows a Michaelis-Menten (MM) function ($MM(x) = x \cdot Vmax/(x + Km)$) with a Vmax of 10 and a Km of 1. The red power law (PL) function ($PL(x) = \alpha \cdot x^g$) was fitted only against the region of the MM function between 0<x<0.3, where it agrees well but diverges for x>0.3. The blue power law was fitted in the region of the Km, 0.7<x<1.3, where it agrees well with the MM function, but diverges everywhere else. The green power law was fitted only against the region of the MM function between 3<x<6, where it agrees well but diverges for x<3. In contrast, the orange power law was fitted against the entire domain of the MM function shown, 0<x<6, and does not appropriately approximate the MM function anywhere except in two points around 1.1 and 4.6. Additionally, above 6 (x>6) all power laws will keep increasing, while the MM function has an asymptotic limit given by Vmax (10 in this example).